\documentclass[english,showpacs,twocolumn,aps,reprint,superscriptaddress]{revtex4-2}
\usepackage{mathptmx}
\usepackage[T1]{fontenc}
\usepackage[latin9]{inputenc}
\usepackage{color}
\usepackage{xcolor}
\usepackage{amsmath}
\usepackage{amssymb}
\usepackage{hyperref}
\usepackage{natbib}
\usepackage{graphics}
\usepackage{graphicx}
\usepackage{epsfig}
\usepackage{bm}
\usepackage{epstopdf}
\usepackage{mathtools}
\usepackage{soul}
\usepackage{xmpmulti}
\usepackage{lipsum}

\usepackage{array}
\usepackage{tabularx}
\usepackage{multirow}
\usepackage[thinlines]{easytable}

\def\vekt#1{\bm{#1}}

\def\vektr{\vekt{r}}

\def\vektx{\vekt{x}}
\def\vekty{\vekt{y}}

\def\vektk{\vekt{k}}

\def\v0{v_0}
\def\I0{I_0}

\def\tp{t{^{\prime}}}

\def\yp{y{^{\prime}}}
\def\tpp{t{^{\prime\prime}}}
\def\ypp{y{^{\prime\prime}}}

\definecolor{dgreen}{rgb}{0.0, 0.5, 0.0}
\definecolor{dblue}{rgb}{0.0, 0.0, 0.5}
\definecolor{green}{rgb}{0.0,0.5,0.0}

\usepackage{babel}

\begin{document}
%
\title{Enhanced terahertz radiation generation by
phase-controlled two-color laser pulses interacting with an under-dense  plasma}

\author{Anjana K P}
\affiliation{Institute for Plasma Research, Bhat, Gandhinagar, 382428, India}
\affiliation{Homi Bhabha National Institute, Training School Complex, Anushaktinagar, Mumbai 400094, India}

\author{Rohit Kumar Srivastav}
\affiliation{Institute for Plasma Research, Bhat, Gandhinagar, 382428, India}

\author{Mrityunjay Kundu}
\email{mkundu@ipr.res.in}

\affiliation{Institute for Plasma Research, Bhat, Gandhinagar, 382428, India}
\affiliation{Homi Bhabha National Institute, Training School Complex, Anushaktinagar, Mumbai 400094, India}

\date{\today}
\begin{abstract}
We investigate terahertz (THz) radiation generation at the vacuum-plasma 
interface driven by the oblique incidence of s-polarized Gaussian laser 
pulse(s) on a 
semi-infinite underdense plasma.~Extending beyond the conventional 
single-frequency bipolar pulse (B-pulse), this 
work focuses on leveraging two-color mixed-frequency pulse-- (M-pulse) 
excitation to enhance THz performance in terms of strength and broadening that 
can be tuned through controlled amplitude and phase of the constituent pulses of 
the M-pulse.
A new expression for the ponderomotive force (PF)---which acts as the main 
driver of THz radiation at the vacuum-plasma boundary---is derived to capture 
the hitherto unexplored effects of phase asymmetry intrinsic to the M-pulse, in 
contrast to a single B-pulse where its phase is irrelevant.
This PF formulation captures the underlying cycle-to-cycle symmetry-breaking for 
the M-pulse field, responsible for efficient THz emission. We demonstrate 
analytically that such M-pulses of the same total energy as a B-pulse may 
generate significantly enhanced PF, leading to THz yields several orders of 
magnitude higher.
With a judicious choice of low-frequency to high-frequency ratio, the M-pulse configuration is shown to emerge as a highly efficient, phase-controllable driver of
 THz radiation and offers a promising route for optimizing THz source design via tailored two-color laser-plasma interactions.
\end{abstract}
\maketitle
\section{Introduction}\label{sec1}
\vspace{-0.3cm}

Terahertz (THz) radiation occupies a unique place in the electromagnetic
spectrum, spanning the range between infrared and microwave regions, primarily 
in the frequency band $\nu_{THz}\approx\! 0.1\!-\!10$~THz~\cite{tonouchi2007cutting,ferguson2002materials}. This spectral 
region has 
historically 
presented significant challenges in terms of source development, {\emph as} 
conventional microwave/infrared technologies struggle to extend into this
domain. Consequently, the development of dedicated THz sources become crucial.

Applications of THz radiation encompass fundamental spectroscopic studies of molecular 
systems~\cite{kang2024time,zhang2017application,pickwell2004simulation,pickwell2004vivo}, 
communications technologies ~\cite{liu2024high,sarieddeen2020next}, 
agricultural applications 
~\cite{li2020study,hadjiloucas1999measurements}, and security 
measures~\cite{tzydynzhapov2020new,palka2012thz}. These varied applications necessitate 
development of efficient, broadband THz sources. While currently available 
commercial sources cover the lower spectral range $\sim 0.3 - 1$~THz,
a substantial technological gap still remains in extending existing 
microwave/infrared sources to generate THz radiation across the entire frequency 
band.

Laser-matter/-plasma interactions provides a promising avenue for
developing high-power, broadband, and tunable tabletop THz 
sources~\cite{johnston2002simulation,dekorsy1996thz,heyman2001terahertz}. Significant
progress has been already achieved in
ultrafast laser technology
and consequently several approaches for THz
sources have emerged. One of
the methods
involves ultrashort laser pulses to excite nonlinear optical effects 
in materials, e.g., optical rectification in electro-optic 
crystals~\cite{carrig1995generation}. Another technique utilizes laser-induced 
plasma in
gas, 
where ponderomotive force drives electron oscillations, resulting in THz 
emission \cite{zhong2006terahertz,xie2007enhancement,thiele2016theory}. The 
laser radiation-pressure may
separate electrons and ions in plasma, thus induces dipole oscillation of electrons
generating THz
radiation~\cite{cheng2001generation}. Short-pulse lasers can excite 
plasma wake-fields, which contain electrostatic oscillations capable of 
mode-converting into electromagnetic (THz) radiation in the presence of plasma 
inhomogeneities/density-gradients~\cite{sheng2005emission,sheng2005powerful}. 
Numerous theoretical/experimental studies have demonstrated that transition 
radiation (TR) 
emitted---when a 
laser pulse traverses the vacuum-plasma interface---can be tuned to the THz 
frequency
range 
under appropriate conditions ~\cite{gorbunov2006transition,frolov2020terahertz,lei2022highly,hua2024enhanced}.
In case of a step-like plasma density, TR is the most effective 
mechanism compared to mode-conversion of induced wake-fields. Enhanced THz 
generation 
through the interaction of laser beat-waves with a plasma
~\cite{malik2012terahertz,bhasin2016laser} and two-color laser excitation 
of air 
plasma have also been demonstrated as an efficient THz source, offering scalability to
high energies~\cite{nguyen2017spectral}. Recent advancements in laser-driven 
semiconductor surfaces have shown promise for generating intense, single-cycle 
THz pulses~\cite{malevich2008thz,si2018terahertz} as well.

Laser-plasma based THz radiation sources in many cases rely on the {\emph 
{ponderomotive force}} (PF) of a shaped laser pulse interacting with an 
underdense 
plasma of appropriate density. The efficacy of this interaction (via PF) directly influences the generated THz radiation in terms of field strength and frequency broadening. Thus it sets our objective investigating various
laser field profiles, including both cycle-to-cycle symmetric (conventional) and 
asymmetric (non-conventional)
pulses, to enhance this PF-driven effects.
Two kinds of pulse asymmetry may be broadly envisaged:
(1) cycle-to-cycle asymmetric pulse and (2) entirely asymmetric pulse.
%
An entirely asymmetric laser-pulse is
characterized by its electric field oscillations between $\vert E_\mathrm{min}\vert$ and $\vert E_0\vert$ (with $\vert E_\mathrm{min}\vert < \vert E_0\vert$), having a
non-zero 
pulse-area~\cite{arkhipov2020unipolar,rosanov2023unipolar,Gorelov_2025} denoted 
by
\vspace{-0.25cm}
\begin{equation}
S = \int_{-\infty}^{\infty} E_L(t) \, dt \neq 0. 
\,\,\,\,\,\,\,\,\,\,\,\,\,\,\,\,\,
\vspace{-0.25cm}
\label{eqSE1}
\end{equation}
When $\vert E_\mathrm{min}\vert = 0$, these pulses \eqref{eqSE1} are purely
unipolar\cite{arkhipov2020unipolar,rosanov2023unipolar}, and 
vertically one-sided. Extremely short 
laser-pulses\cite{Calegari_2016,Hassan_2016,zeng2007generation,
jones1993ionization,you1995generation}, e.g, 
atto-seconds pulses, 
half-cycle or near single-cycle pulses often satisfy $S\!\ne0$.
When $S\!\!=\!\!0$, the laser pulse is up-down symmetric (bipolar).~Certain
class of bipolar pulses may have left-right asymmetry w.r.t. the pulse 
center (e.g., laser pulses steepened at the leading edge or at 
the trailing 
edge\cite{rousseau2006efficient,matsuoka2008pulse,Gopal_2016,SINGH_2016}), which 
are not considered here.~The 
inequality~\eqref{eqSE1}
ensures that charged particles receive a non-zero 
average 
momentum {\emph{after the pulse end}}.
In this work, on the other hand, we propose cycle-to-cycle asymmetric laser 
pulses where up-down 
symmetry between two successive laser-cycles or half-cycles is broken and 
locally satisfy
\vspace{-0.15cm}
\begin{equation}
S = \int_{t_c - l T_0/2}^{t_c + l T_0/2} E_L(t) \, dt \neq 0.
\label{eqSE}
\end{equation}
Where, $t_c$ is the center of a cycle, $l\!=\!1,2,3\cdots$ is an integer, 
$T_0$ is the 
(approximate) time-period about $t_c$. The condition~\eqref{eqSE} may be
recognized as a general form of~\eqref{eqSE1}. But the electric
field $E_L(t)$ in \eqref{eqSE} may still satisfy $\int_{-\infty}^{\infty} E_L(t)
\, dt = 0$ {\emph{globally}} in some cases. Such {\emph{local symmetry
breaking}} may be possible 
by mixing two (or more) pulses of different frequencies with controlled phases,
and it 
may deliver non-zero unidirectional average momentum to charged particles 
{\emph{locally}}. At high laser intensities, due to non-linear laser-plasma 
interactions, 
a substantial part of this locally acquired momentum may be retained by the 
charged particles after the laser pulse (similar to the case of laser-pulse 
steepening).
However, the impact of this phenomenon
on the ponderomotive force exerted
to an underdense-plasma and subsequently on THz radiation emissions, remains
largely unexplored.

The present work aims to the generation of THz radiation via TR at the 
vacuum-plasma
boundary resulting from the oblique incidence of two mixed frequency (thus 
two-color) Gaussian laser pulses 
(M-pulse) on a semi-infinite underdense plasma. The constituent pulses of the 
M-pulse may have different amplitude and phases; thus it may
satisfy either $S=0$, or $S\ne 0$ (unipolar case). In addition to the
mostly used conventional {\emph{single}} Gaussian bipolar laser-pulse (B-pulse)
~\cite{frolov2020terahertz}, we examine the effect of M-pulses
with wider frequency combinations to the PF generation, 
and subsequent THz
radiation. As the generated PF serves as the source of THz radiation in this 
case, {\emph{a new form}} of PF has been derived that depends mainly on the 
different phase-combinations of the constituent pulses of the M-pulse, otherwise 
phase plays no role for a {\emph{single}} B-pulse. A {\emph{phase-optimized, 
hitherto unexplored}} PF with a M-pulse of same energy of a {\emph{single}} 
B-pulse, is shown to produce {\emph{orders of magnitude}} stronger PF and 
corresponding THz radiation. The new PF expression connects 
wider range of low to high frequency ratio ($0<r<1 $) for the two-color 
phase-optimized M-pulse, thus covering unipolar regime (not discussed here) 
to the beating regime. By shortening the input laser pulse-duration, the 
spectral-width $(\Delta\omega)$ of the PF and the corresponding THz 
radiation-spectrum broadens, and their peaks shift towards higher 
radiation frequency $\omega$.

%

The phase-dependent PF with M-pulses may have wider impact in other research areas, particularly laser-wakefield acceleration of electrons, ion-acceleration and laser-self focusing in under-dense plasmas.

The article is arranged beginning with the B-pulse/M-pulse laser interaction 
with an under-dense plasma and THz generation process via 
transition radiation in Sec.\ref{sec2}. The PF expression for the phase-dependent 
M-pulse and its optimization is given in Sec.\ref{sec3}. Results are 
given in Secs.\ref{sec4} and Sec.\ref{sec5} followed by summary in 
Sec.\ref{sec6}.

\vspace{-0.35cm}

\section{Laser interaction at a plasma surface}\label{sec2}
\vspace{-0.25cm}
\begin{figure}[h]
\vspace{-0.5cm}
\includegraphics[width=0.5\linewidth]{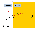}
\vspace{-0.25cm}
\caption{\hspace{-0.1cm}Schematic of s-polarized laser pulse interaction with an
under-dense plasma at the vacuum-plasma boundary. A single laser pulse or two laser pulses may be incident at an angle $\alpha$ in the $xy$-plane of incidence. Here, $y',y'', \tilde{y}$
represent the incident, reflected and the transmitted field directions respectively.}
\vspace{-0.25cm}
\label{Figure1_label}
\end{figure}

Let us consider oblique incidence of two s-polarised Gaussian laser pulses 
($j=1,2$) on a
step-like semi-infinite vacuum-plasma interface {{(see schematic 
in Fig.\ref{Figure1_label})}} at $y=0$; where $+y$ is the direction of 
propagation of a laser in case of normal incidence. These pulses are 
characterised by individual field amplitude $E_j$, frequency $\omega_j 
= 2\pi 
c/\lambda_j$, wavelength $\lambda_j$, and the pulse duration $\tau$; 
where $c$ is the speed of light in vacuum. For convenience, it is assumed that 
each laser pulse is relatively long so that $\omega_j \tau \gg 1$ or $\tau$ is 
many times the laser period $\tau_j=2\pi/\omega_j$. Each pulse strikes the 
vacuum-plasma interface (from left) at an angle $\alpha_j$ relative to the 
surface-normal to the plasma. The plasma is assumed to be underdense (for both 
$\omega_1, \omega_2>\omega_1$) with 
frequency $\omega_p = \sqrt{e^2 N_e/m_e \epsilon_0}$,
ion density $N_0$, and electron density $ N_e(y) = N_{0}\Theta(y)$;
where
\vspace{-0.25cm}
\begin{equation}
\begin{aligned}
\Theta(y) = \begin{cases} 
0, & y < 0 \,\,\,\,\,\, (vacuum)\\
1, & y \geq 0 \,\,\,\,\,\, (plasma)
\end{cases}
\end{aligned}
\vspace{-0.25cm}
\label{eq2}
\end{equation}
is the Heaviside unit-step function.

The total electric field of the jointly incident s-polarized laser pulses 
(M-pulse) may 
be written as
\vspace{-0.25cm}
\setlength{\jot}{-5.0pt}
\begin{align}
\nonumber
%
\vekt{E}_{L} 
\!\!= \vekt{\hat{z}}({E_{0}}/{2})
\exp\left[-{(\tp \!\!-t_0)^2}/{2\tau^2}\right]\!\! \sum_{j=1}^{2}\!\! a_j
\exp\left[{-i(\omega_j 
\tp+ \beta_j)}\right] \\
& \hspace{-8.0cm} + \,\, c.c.
\label{eq3}
\end{align}
\setlength{\jot}{0pt} 
\vskip -0.25cm
\noindent
We assume both the constituent pulses have same duration~$\tau$, angle of
incidence $\alpha_j\!\! =\!\! \alpha$, and pulse centers at $t_0$ for the simplicity of
analysis, and same integral area under their envelopes. However, we 
introduce the additional phase $\beta_j$ in the respective carrier and write 
$E_j = a_j E_0$, whose importance will be apparent in the subsequent sections.
Here {{$\tp = t - \yp/c$}} is the retarded time, $\yp =
y\cos{\alpha} + 
x\sin{\alpha}$ is the oblique co-ordinate in the $xy$-plane of incidence along the 
incident light propagation $\hat{\vektk}_I = \hat{\vekty}\cos{\alpha} + 
\hat{\vektx}\sin{\alpha}$ so that $\yp = \hat{\vektk}_I. \vektr$. 
Setting $a_1=0, \, a_2=1, \, \beta_1 = 0, \, \beta_2 = 0$, the commonly used 
Gaussian B-pulse~\cite{frolov2020terahertz} is readily found as
\vspace{-0.15cm}
\begin{equation}
\vekt{E}_{L} = \vekt{\hat{z}}(E_{0}/2) 
\exp\left(-{(\tp-t_0)^2}/{2\tau^2}\right) \exp{(
-i\omega_2 \tp)} + c.c.
\label{eq4}
\end{equation}
The coefficients $a_j$ are chosen in such a way that the
energy of the M-pulse \eqref{eq3} is same 
as the 
energy of the B-pulse~\eqref{eq4}.This allows to compare 
the effects of B-pulse and the M-pulse of same energy on the 
plasma electron oscillation and low-frequency THz emission as well. 
%
The total laser field in the vacuum $(y\leq 0)$ for the M-pulse 
\eqref{eq3} is obtained as
%
\begin{align}
\vekt{E}_{L}^{R}(\bold{r}, t) &= \bold{\hat{z}} \tau \sqrt{2\pi} 
\frac{E_{0}}{2}  
\int_{-\infty}^\infty \frac{d\omega}{2\pi}  
\bigg[ \sum_{j=1}^{2} a_j \exp{\left(\!\!-\frac{(\omega-\omega_j)^2 
\tau^2}{2}\right)} \nonumber \\
&\quad \hspace{-1cm}
\exp\left[{i(\omega - \omega_j)t_0 + i \beta_j}\right]
\bigg] \exp{(-i\omega \tp)} \nonumber \\ 
&\quad\hspace{-1cm} + \bold{\hat{z}}  \tau \sqrt{2\pi}
\frac{E_{0}}{2}  
\int_{-\infty}^\infty \frac{d\omega}{2\pi}  
\bigg[  \sum_{j=1}^{2} R_j a_j \exp{\left(\!\!-\frac{(\omega-\omega_j)^2 
\tau^2}{2}\right)} \nonumber \\
&\quad\hspace{-1cm} 
\exp\left[{i(\omega - \omega_j)t_0 + i \beta_j}\right]
\bigg] \exp{(-i\omega \tpp)} 
 + c.c., \hspace{0.25cm} y \leq 0
\label{eq5}
\end{align}
%
where {{$\tpp = t  - \ypp/c$}}, $\ypp =
(x\sin{\alpha}-y\cos{\alpha})$.
%
Corresponding total field transmitted into the plasma 
region $(y \geq 0)$ reads
\vspace{-0.25cm}
\begin{align}
\vekt{E}_{L}^{T}(\bold{r}, t) &= \bold{\hat{z}} \tau \sqrt{2\pi} 
\frac{E_{0}}{2}  
\int_{-\infty}^\infty \frac{d\omega}{2\pi}  
 \bigg[ \sum_{j=1}^{2} T_j a_j  \exp{\left(\!\!-\frac{(\omega-\omega_j)^2 
\tau^2}{2}\right)} \nonumber \\
&\quad \hspace{-1.5cm}
\exp\left[i(\omega - \omega_j)t_0 + i\beta_j\right]
\bigg] \exp{(-i\omega \tilde{t}(\omega))} + c.c.,\hspace{0.25cm} y \geq 0
\label{eq6}
\end{align}
where {{$\tilde{t}(\omega) =
t-\tilde{y}/c$,
$\tilde{y} =
\left(x\sin{\alpha}+y\sqrt{\epsilon(\omega)-\sin^2{\alpha}}\right)$}}, $\epsilon(\omega) = 1-{\omega_p^2}/{\omega^2}$.
%
Equations \eqref{eq5} and \eqref{eq6} include the reflection coefficient $R_j = R_j(\omega,\alpha)\!\! =\!\! (C - S)/(C + S)$, and transmission 
coefficient $T_j \!\! =\!\! T_j(\omega,\alpha) \!\! =\!\! 2 C/(C + S)$ for s-polarized light
%
%
with $C = \cos{\alpha}$ and $S = \sqrt{\epsilon(\omega)-{\sin^2{\alpha}}}$.
Under the condition $\omega_j\tau\gg 1$, $R_j(\omega,\alpha)$ and 
$T_j(\omega,\alpha)$ varies slowly within the bandwidth $\Delta\omega$; and can 
be approximated as $T_j(\omega,\alpha) = T_j(\omega_j,\alpha)$ and $
R_j(\omega,\alpha) = R_j(\omega_j,\alpha)$ for $j=1,2$ respectively. It permits
$T_j, R_j$ out of the integrals~\eqref{eq5}-\eqref{eq6}, and their 
evaluation analytically\cite{frolov2020terahertz}.

\vspace{-0.35cm}
\subsection{Excitation of THz fields}
\vspace{-0.25cm}
The low-frequency ($\omega$) terahertz emission into the
vacuum may be investigated by linearized the Maxwell's equations for the fields 
($\vekt{E}, \vekt{B}$) and the electron
equation of motion for the
hydrodynamic velocity $\vekt{v}$. All the dominant high frequency
quantities $f_0(\omega_j)$  are separated as $f=f_0 + f^1$.
The low-frequency $f^1(\omega)$, i.e.,
$\vekt{E}^1, \vekt{B}^1, \vekt{v}^1$ satisfy following equations (without 
superscript)
\begin{eqnarray}
\vekt{\nabla}\times \vekt{E}(\bold{r},t) & = & 
-\frac{1}{c}\frac{\partial\vekt{B}(\bold{r},t)}{\partial t}
\label{eq8}
\\
\vekt{\nabla}\times \vekt{B}(\bold{r},t) & = & \frac{4
\pi}{c}eN_e(y)\vekt{v}+\frac{1}{c}\frac{\partial\vekt{E}(\bold{r},t)}{\partial 
t} 
\label{eq9}
\\
\frac{\partial \vekt{v}(\bold{r},t)}{\partial t} & = & \frac{e}{m_e} 
\left[\vekt{E}(\bold{r},t) - \vekt{\nabla} \phi(\bold{r},t) \right]
\label{eq10}
\end{eqnarray}
where $\vekt{F}_p = - {e} \vekt{\nabla }\phi(\bold{r},t)$ is
the PF exerted by the laser pulse on the plasma electrons, 
and 
$\phi(\bold{r},t)$ is the corresponding ponderomotive potential (PP). Since PF 
lies in the $xy-$plane of incidence, the generated low
frequency fields $\vekt{E}, \vekt{B}$ are always p-polarized (with $B_z$, $E_x$, $E_y$).
%
%
%
Equations \eqref{eq8}-\eqref{eq10} are Fourier transformed w.r.t. time, giving
%
%
\setlength{\jot}{4.0pt}
\begin{align}
& \quad\hspace{-0.65cm}\frac{\partial}{\partial y}\left[\frac{1}{\epsilon(\omega,y)}\frac{\partial
B_z(\bold{r},\omega)}{\partial y}\right] +
\frac{\omega^2}{c^2}\left[1-\frac{\sin^2{\alpha}}{\epsilon(\omega,y)}\right]
B_z(\bold{r},\omega) = \nonumber \\
& \quad\hspace{-0.65cm}\frac{\omega^2}{c^2}\sin{\alpha}\left\lbrace
\frac{d}{dy}\!\!\left[\frac{\phi(\bold{r},\omega)\omega_p(y)^2}{\omega^2\epsilon(\omega,y)} \right]\!\! - \!
\frac{\omega_p(y)^2}{\omega^2\epsilon(\omega,y)}\frac{d\phi(\bold{r},\omega)}{dy}
\right\rbrace
\label{eq11}
\end{align}
where $\epsilon(\omega,y) = 1-{\omega_p^2(y)}/{\omega^2}$, and 
$\phi(\bold{r},\omega)$ is the Fourier transformed PP. Equation~\eqref{eq11} 
determines low-frequency $B_z$ in the plasma and vacuum, with appropriate 
boundary conditions at the vacuum-plasma interface $y=0$. Its solution is 
contingent upon the frequency-domain $\phi(x,y=0, \omega)$ at the 
interface $y=0$, and can be expressed as
\setlength{\jot}{0.0pt}
\begin{align}
B_z(\bold{r},\omega) &= \frac{i\omega_p^2(y)\sin{\alpha}}{c\omega}
\frac{\phi(x,y=0,\omega)}{\epsilon(\omega,y)\cos{\alpha}  
+ \sqrt{\epsilon(\omega,y)-\sin^2{\alpha}}} \nonumber \\  
&\quad \times \exp{\left(-\frac{i\omega}{c}y\cos{\alpha} \right)}, \quad y \leq 0.
\label{eq12}
\end{align}

\vspace{-0.25cm}
\begin{align}
B_z(\bold{r},\omega) &= \frac{i\omega_p^2(y)\sin{\alpha}}{c\omega}
\frac{\phi(x,y=0,\omega)}{\epsilon(\omega,y)\cos{\alpha}  
+ \sqrt{\epsilon(\omega,y)-\sin^2{\alpha}}} \nonumber \\  
&\quad \times \exp{\left(\frac{i\omega}{c}y\sqrt{\epsilon(\omega,y)-\sin^2{\alpha}} \right)}, 
\quad y \geq 0.
\label{eq13}
\end{align}
At a normal incidence, $B_z(\bold{r},\omega)$ becomes zero, and no new
frequency ($\omega$) radiation is emitted.
Importantly, the strength and the profile of emitted low-frequency field $B_z(\bold{r},\omega)$
directly depends on the structure and strength of  
$\phi(\bold{r},\omega)$. Therefore, we look for maximizing the PF/PP by varying laser parameters, while keeping the input pulse energy fixed. This is calculated for two distinct scenarios: (i) an
input M-pulse and (ii) an input B-pulse, as described in Sec.\ref{sec3}.

We choose the laser wavelength $\lambda_2=800$~nm, frequency $\omega_2 = 2\pi 
c/\lambda_2 = 2.35\times 10^{15}$~rad/sec, field strength $E_0 = 2.5711 
\times 10^{11}$~V/m and the plasma frequency $\omega_p = 0.07\omega_2$ ($f_p
=\omega_p/2\pi\approx 26.18 THz$) unless mentioned explicitly, otherwise.
\vspace{-0.25cm}
\section{Calculation of Ponderomotive Force}\label{sec3}
\vspace{-0.25cm}
In the non-relativistic case, the equation of motion of an electron in the oscillating electric and magnetic fields ($\vekt{E}_L, \vekt{B}_L$) of an input laser pulse is given by
\begin{equation}
m_e\frac{d\bold{v}(\bold{r},t)}{dt} = e\left[\vekt{E}_L(\bold{r},t) 
+ \bold{v}(\bold{r},t)\times \vekt{B}_L(\bold{r},t)/c \right].
\label{eq14}
\end{equation}
Equation~\eqref{eq14}
signifies two kinds of non-linear motion: (i) evaluation of 
$\vekt{E}_L(\bold{r},t)$ at the electron's actual position~$\bold{r}$ due to 
the field's dependence on its changing position, creating a feedback loop where 
the electron motion influences the field and vice versa; and (ii) second-order
nonlinear coupling $\bold{v}(\bold{r},t)\times\vekt{B}_L(\bold{r},t)$. These 
non-linearities are the source of PF.
\vspace{-0.5cm}
\subsubsection{Ponderomotive force for mixed-frequency laser pulses}
\vspace{-0.25cm}
We follow F.F. Chen~\cite{chen1984introduction} for the PF/PP. Consider a two-color mixed
laser field $\vekt{E}_L(\bold{r},t) = 
\vekt{E}_s(\bold{r})\sum_{j=1}^{2}a_j\cos{(\omega_j t+\beta_j)}$, where
$\vekt{E}_s(\bold{r})$ represents spatial part with $a_j$ as a coefficient. This space-time separation
may not be possible in some cases, and $\vekt{E}_s$ may include 
time-varying envelopes, i.e., $\vekt{E}_s(\bold{r},t)$ in general. But due to analytical tractability, this space-time separation procedure is
{\emph{conventionally followed}}.
First disregard $\bold{v}(\bold{r},t)\times\vekt{B}_L(\bold{r},t)$, and 
determine $\vekt{E}_L(\bold{r}_0,t)$ at the electron's initial position 
$\bold{r}_0$. Using ${d\bold{v}_1}/{dt} = 
e\vekt{E}_L(\bold{r}_0,t)/m_e$, one 
may assess the first-order velocity $\bold{v}_1$ and displacement $\bold{r}_1$ 
as
\vspace{-0.2cm}
\begin{eqnarray}
\begin{aligned}
%
\bold{v}_1 & = \frac{e}{m_e} 
\vekt{E}_s\sum_{j=1}^{2}\frac{a_j}{\omega_j}\sin{(\omega_j t+\beta_j)}\\
%
\bold{r}_1 & = 
- \frac{e}{m_e}\vekt{E}_s\sum_{j=1}^{2}\frac{a_j}{\omega_j^2}\cos{(\omega_j 
t+\beta_j)}.
\end{aligned}
\label{eq17}
\vspace{-0.2cm}
\end{eqnarray}
When $\vekt{E}_s$ varies with time, solutions \eqref{eq17} for
$\bold{v}_1, \bold{r}_1$ are invalid. We may use slowly varying envelope 
approximation, and replace $\vekt{E}_s(\bold{r}) 
\rightarrow\vekt{E}_s(\bold{r},t)$ in \eqref{eq17} to obtain approximate 
$\bold{v}_1, \bold{r}_1$.
Taking the Taylor expansion $
\vekt{E}_L(\bold{r},t) = \vekt{E}_L(\bold{r}_0,t)  + 
(\bold{r}_1.\vekt{\nabla)}\vekt{E}_L\vert_{\bold{r}=\bold{r}_0} + \cdots$ about 
$\bold{r}_0$,
and the effect of $(\bold{v}_1\times\vekt{B}_1)$ with $
\vekt{B}_1 = -c\vekt{\nabla}\times\vekt{E}_s\sum_{j=1}^{2}a_j \sin{(\omega_j 
t+\beta_j)}/\omega_j$, the second-order component of force \eqref{eq14} on 
the electron
%
is found as
\vspace{-0.25cm}
\begin{equation}
m_e\frac{d\bold{v}_2}{dt} = 
e\left[(\bold{r}_1.\vekt{\nabla})\vekt{E}_L + 
\bold{v}_1\times \vekt{B}_1 / c \right].
\label{eq21}
\vspace{-0.2cm}
\end{equation} 
%
Equation~\eqref{eq21} may be regarded as the PF on the electron. One clearly 
finds from \eqref{eq21} two kinds (terms) of non-linearities, as discussed above.
Using ($\bold{r}_1, \bold{v}_1, \vekt{B}_1$) we explicitly write
\vspace{-0.2cm}
%
\begin{align}
\nonumber
\frac{d\bold{v_2}}{dt} = -\frac{e^2}{m_e} \Bigg[ \sum_{i=1}^{2} \sum_{j=1}^{2}
\frac{a_i a_j}{\omega_i^2} \cos{(\omega_i t+\beta_i)}
\cos{(\omega_j
t+\beta_j)} \\ \nonumber
&\quad\hspace{-8.0cm} \times
(\vekt{E_s} \cdot \vekt{\nabla})\vekt{E_s}
 + \sum_{i=1}^{2} \sum_{j=1}^{2} \frac{a_i a_j}{\omega_i 
\omega_j} \sin{(\omega_i t+\beta_i)} \sin{(\omega_j t+\beta_j)} \\
&\quad\hspace{-8.0cm} \times \vekt{E_s} 
\times (\vekt{\nabla} \times \vekt{E_s}) 
\Bigg].
\label{eq22}
\vspace{-0.2cm}
\end{align}
For the s-polarized pulses here, $(\vekt{E}_s \cdot
\vekt{\nabla})\vekt{E}_s = \vekt{0}$.
The average PF is {\emph{often conventionally}} obtained from the time-average of these 
nonlinear interaction terms over {\emph{one-period}} (to eliminate high-frequency 
variation) in case of a monochromatic laser field.
However, in the case of multi-color or two-color laser field (as the M-pulse), 
this
time-average is {\emph{weakly justified}}. Making an attempt to 
average over the high-frequency ($\omega_2$), leaves behind terms of 
low-frequency ($\omega_1$) of the input laser. This yields average PF different 
from the {\emph{conventional case}}, which is also true for a broadband input 
laser pulse.
Due to frequency mixing ($\omega_2, \omega_1$), we
average $m_e{d\bold{v}_2}/{dt}$ on the higher frequency ($\omega_2$)
oscillations, leading to
%
\begin{equation}
\begin{aligned}
&\quad\hspace{-3.5cm}
\vekt{F}_p(\bold{r},t) = \left< m_e \frac{d\bold{v}_2}{dt} \right>_{\tau_2}
 \approx
-\frac{e^2 \vekt{\nabla} E_s^2}{4m\omega_2^2} \Psi 
\end{aligned}
\label{eq23a}
\end{equation}
\vspace{-0.5cm}
\begin{equation}
\begin{aligned}
&\quad\hspace{-0.6cm} \text{and} \hspace{0.25cm}
\Psi =
a_2^2 + \frac{a_1^2}{\omega_{12}^2} \Bigg[
1 - \frac{\sin{(4\pi\omega_{12} + 2\beta_1)}}{4\pi\omega_{12}} 
+ \frac{\sin{2\beta_1}}{4\pi\omega_{12}} 
\Bigg] \\
&\quad\hspace{-0.7cm} + \frac{a_1 a_2}{\pi\omega_{12}} \Bigg[
\frac{\sin{(2\pi - 2\pi\omega_{12} + \beta_2 - \beta_1)}}{1-\omega_{12}} 
- \frac{\sin{(\beta_2 - \beta_1)}}{1-\omega_{12}} \\
&\quad\hspace{-0.7cm} - \frac{\sin{(2\pi\omega_{12} + \beta_1 +
\beta_2)}}{1+\omega_{12}} 
+ \frac{\sin{(\beta_1 + \beta_2)}}{1+\omega_{12}} 
\Bigg]
\end{aligned}
\label{eq23}
\vspace{-0.2cm}
\end{equation}
where $\omega_{12}\!\!=\!\!\omega_1/\omega_2$.~While time-averaging, the 
integration over the slowly varying envelope 
$\vekt{E}_s(\bold{r},t)$ is {\emph{customarily suppressed}}. The corresponding 
$\phi(\bold{r},t)$ is 
generated by
$\vekt{F}_p(\bold{r},t)\!\! =\!\! - e\vekt{\nabla}\phi(\bold{r},t)$, or 
$\vekt{F}_p(\bold{r},\omega)\!\! = \!\! - 
e\vekt{\nabla}\phi(\bold{r},\omega)$.~The {\emph{new}} 
function 
$\Psi(a_1, a_2, \beta_1, \beta_2, \omega_1, \omega_2)$ in \eqref{eq23} is 
defined for convenience, giving
\vspace{-0.25cm}
%
\begin{equation}
\phi(\bold{r},t) =
\frac{e E_s^2(\bold{r},t)}{4m\omega_2^2} \Psi .
\label{eq25}
\vspace{-0.2cm}
\end{equation}
For $a_1=0, a2=1, \beta_1=\beta_2=0$, the input M-pulse reduces to the 
conventional monochromatic B-pulse as given by \eqref{eq4} and the PP gets simplified (as 
$\Psi\rightarrow 1 $) to the welknown form
\vspace{-0.25cm}
\begin{equation}
\phi(\bold{r},t) =  -\frac{e E_s^2(\bold{r},t)}{4m\omega_2^2}.
\label{eq26}
\vspace{-0.25cm}
\end{equation}

Eventually, using the transmitted field \eqref{eq6}  into the plasma
region $(y \geq 0)$ in \eqref{eq25} and \eqref{eq26},
 corresponding $\phi(\bold{r},t) $ at the vacuum-plasma interface $y=0$ are 
calculated for the M-pulse ($\phi_M$) and the B-pulse ($\phi_B$) in the time and 
frequency domain (making $a_j \rightarrow a_j \vert T_j(\omega_j,\alpha)\vert$, 
$\Psi(a_1, a_2, \beta_1, \beta_2, \omega_1, \omega_2)\!  \rightarrow \! \Psi(a_1 
T_1, a_2 T_2, \beta_1, \beta_2, \omega_1, \omega_2, \alpha)$ as
 %
%
\vspace{-0.1cm}
\begin{eqnarray}
\begin{aligned}
& \quad\hspace{-40pt}\phi_M(x,y=0,t)  = \frac{\Psi}{|T_2(\omega_2,\alpha)|^2} 
\,\,\, \phi_B(x,y=0,t) \\
\label{eq27a}
& \quad\hspace{-40pt}\phi_M(x,y=0,\omega)  = 
\frac{\Psi}{|T_2(\omega_2,\alpha)|^2} \,\,\, \phi_B(x,y=0,\omega)
\end{aligned}
\label{eq28a}
\end{eqnarray}
\vspace{-0.5cm}
\begin{eqnarray}
\begin{aligned}
& \quad\hspace{-28pt} \phi_B(x,y=0,t) \\ &\quad\hspace{-28pt}=  \phi_0 
|T_2(\omega_2,\alpha)|^2\exp{\left[-\frac{1}{\tau^2}\left(t-t_0-\frac{x}{c}\sin{\alpha} \right)^2 \right]}\\
\label{eq29}
& \quad\hspace{-28pt}
\phi_B(x,y=0,\omega) \\ &\quad \hspace{-28pt}=  \sqrt{\pi}\tau 
\phi_0 
|T_2(\omega_2,\alpha)|^2\exp{\left(i\frac{\omega}{c}x\sin{\alpha} - 
\frac{\omega^2\tau^2}{4} +i\omega t_0\right)}
\end{aligned}
\label{eq30}
\end{eqnarray}
where $\phi_0={eE_{0}^2}/{4m_e\omega_2^2}$. The expressions of PP~\eqref{eq30}
for the monochromatic B-pulse follows Ref.~\cite{frolov2020terahertz}, and it {\emph {does 
not}} depend on the light phase. However, the PP for the M-pulse~\eqref{eq28a},
clearly shows additional nonlinear coupling due to superposition of different 
light phases $\beta_1, \beta_2$, frequency ratio 
$\omega_{12}=\omega_1/\omega_2$, transmission coefficients $T_1, T_2$, and 
amplitude coefficients $a_1, a_2$ through $\Psi$. The assumption we made for the M-pulse was
that the constituent pulses have a common slowly varying envelope (last factor 
in \eqref{eq30}) as the B-pulse. Otherwise, calculation of the PF/PP would have been more involved, and in some cases may not be possible analytically.

%

\vspace{-0.25cm}
\subsection{Optimization of the Ponderomotive force}
\vspace{-0.25cm}
As the generated low-frequency radiation field [from \eqref{eq12} and 
\eqref{eq13}] linearly depends on the respective PP ($\phi_M$ for the M-pulse), 
it is {\emph{important}} to maximize $\Psi$ in \eqref{eq23} to yield the 
maximum 
output field strength.
The form of $\Psi$ suggests that a judicious choice of parameters $(a_1, a_2, 
\beta_1, \beta_2, \omega_1, \omega_2, \alpha)$, {\emph{more importantly}} the 
phases $\beta_1, \beta_2$ can maximize the PP and the associated PF through 
$\Psi$. We fix ($\omega_2, \alpha$) and set ratio $\omega_{12}$. The 
coefficients $a_1, a_2$ are dynamically chosen such that energy of 
the M-pulse equals the energy of the B-pulse. With these constraints, we search 
for phases $-2\pi \leq \beta_1, \beta_2 \leq 2\pi$ to maximize/minimize $\Psi(a_1, a_2, 
\beta_1, \beta_2, \omega_1, \omega_2, \alpha)\rightarrow \Psi(\beta_1, \beta_2)$. 
 {{Four distinct $\omega_{12} = 0.1, 0.2, 0.3, 0.9$}} are chosen for
illustration,
with $\alpha=\pi/4$, and $\omega_2$ corresponding to $800$~nm laser 
wavelength. For each $\omega_1$ (or $\omega_{12}$),
various combinations of $\beta_1, \beta_2$ (at least thirty) have been tried 
numerically, to obtain the surface of $\Psi(\beta_1, \beta_2)$. 
\begin{figure}[hbt!]
        \centering
\includegraphics[width=0.235\textwidth]{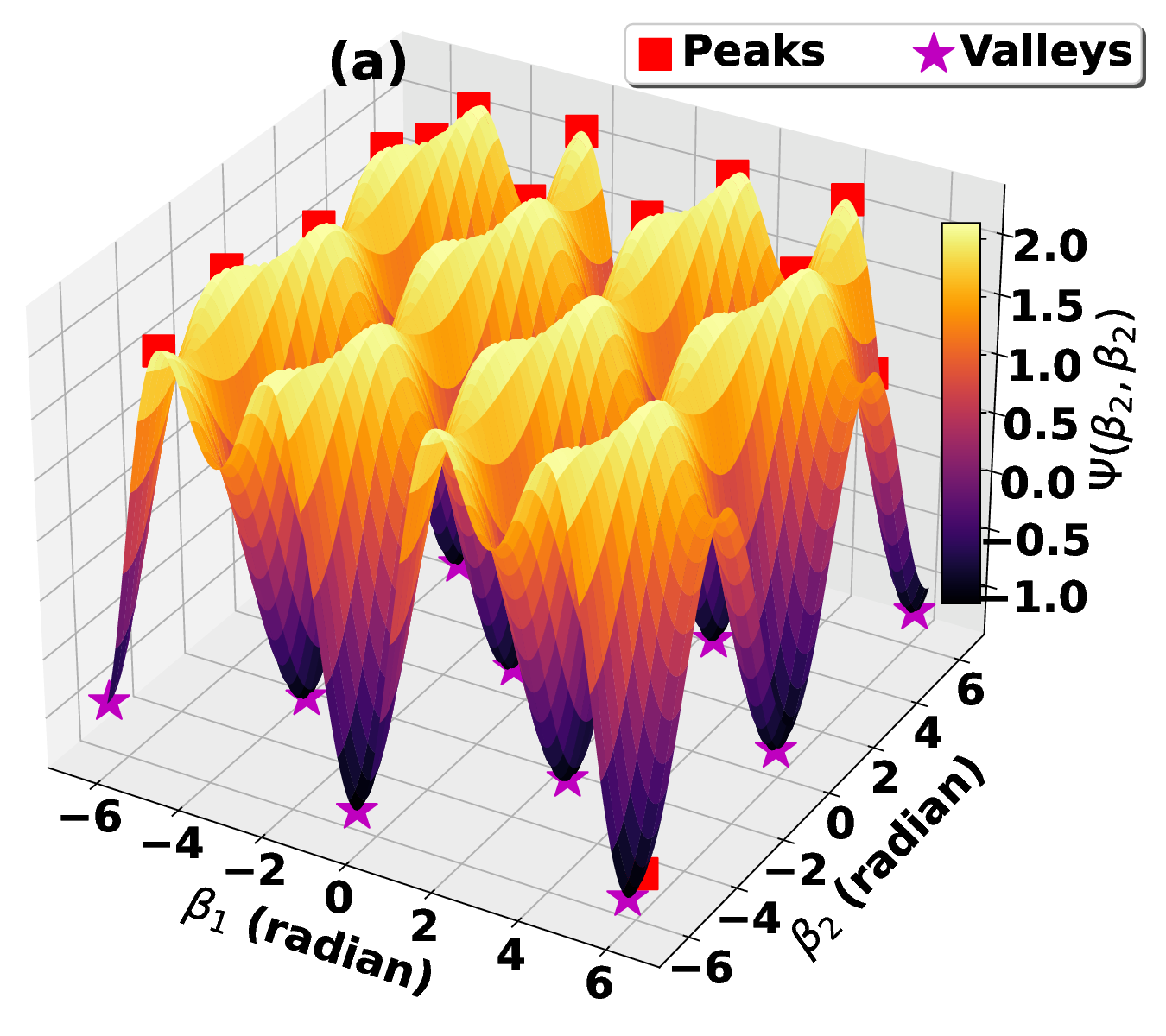}
\includegraphics[width=0.235\textwidth]{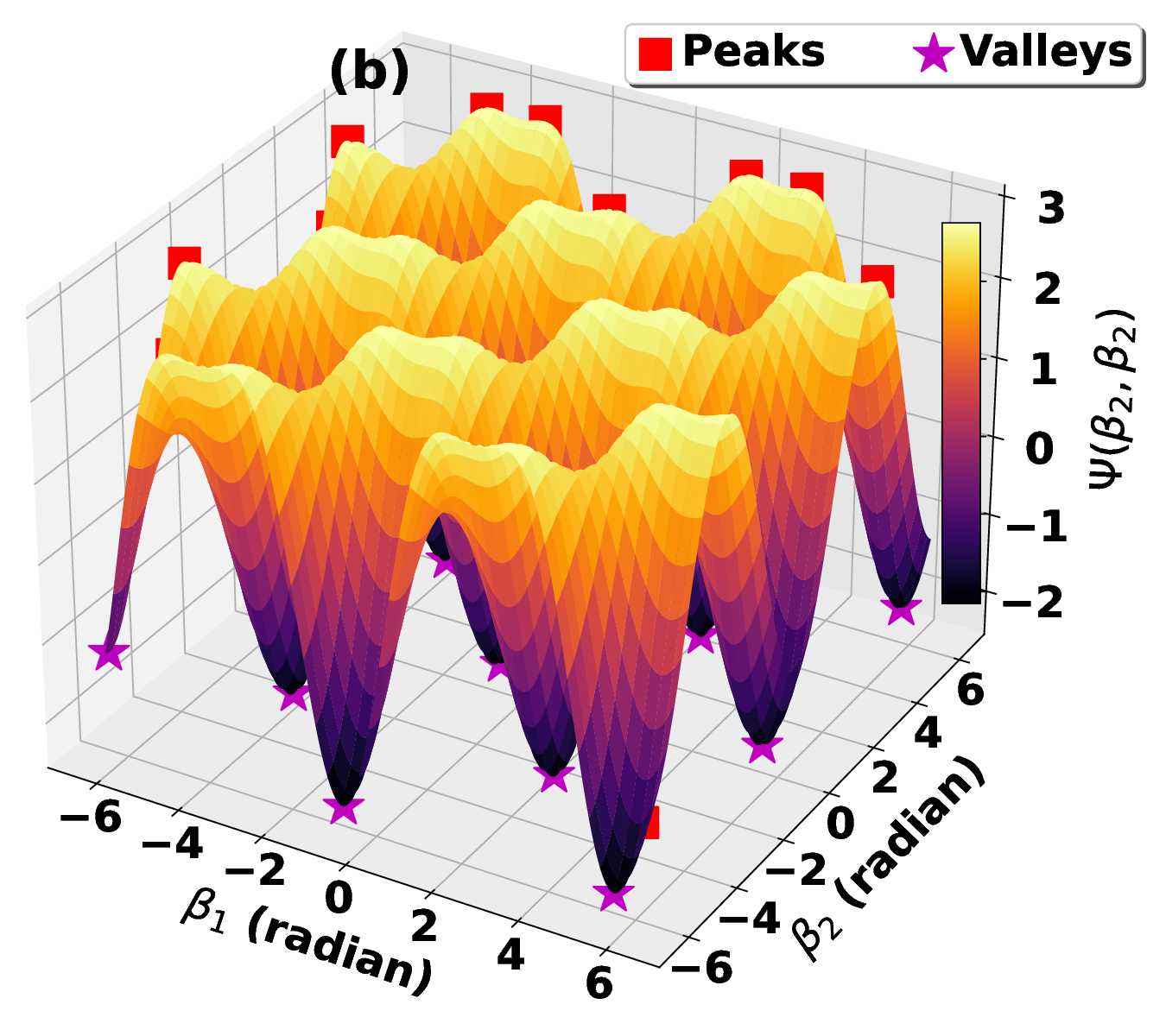}
\includegraphics[width=0.235\textwidth]{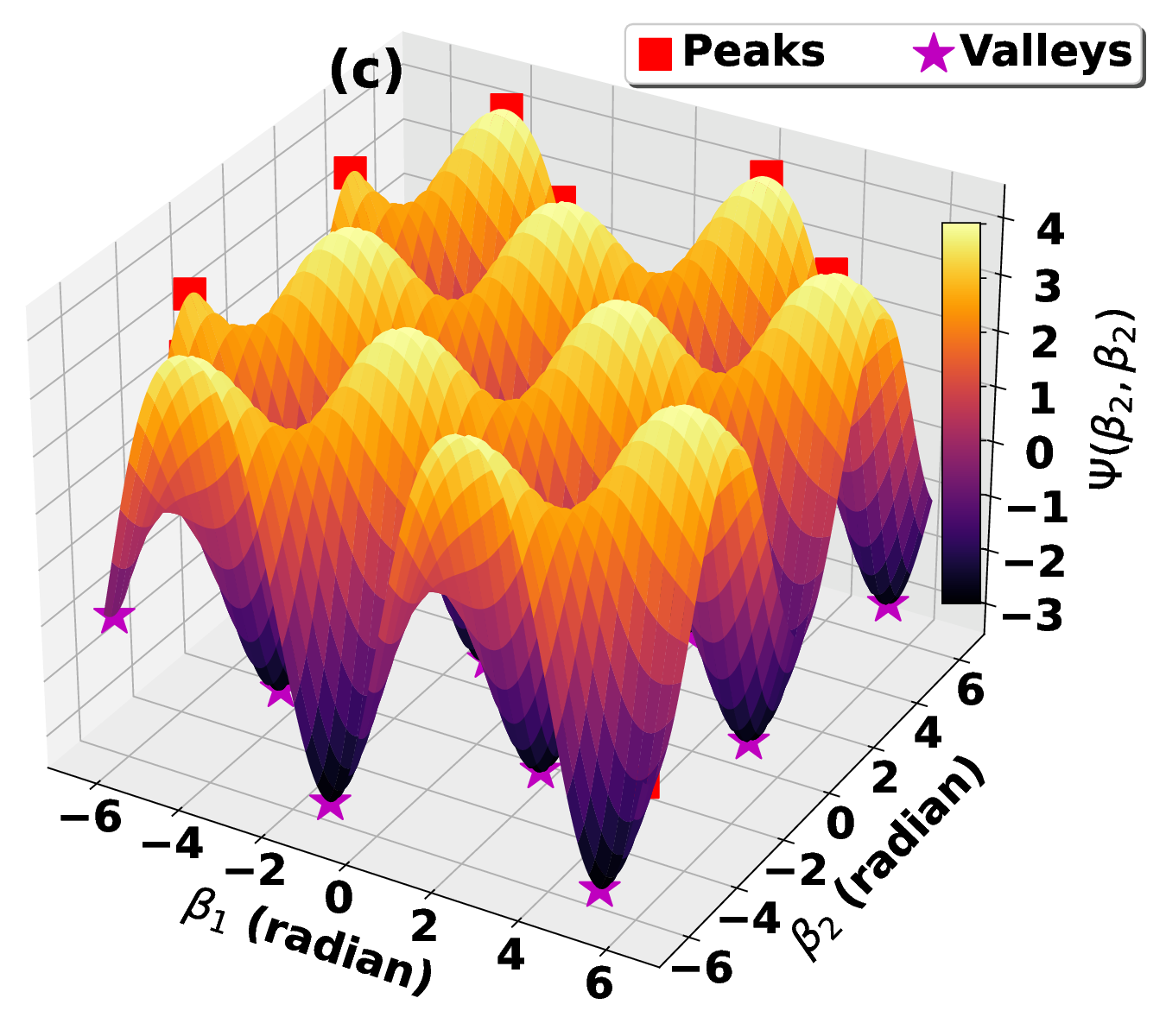}
\includegraphics[width=0.235\textwidth]{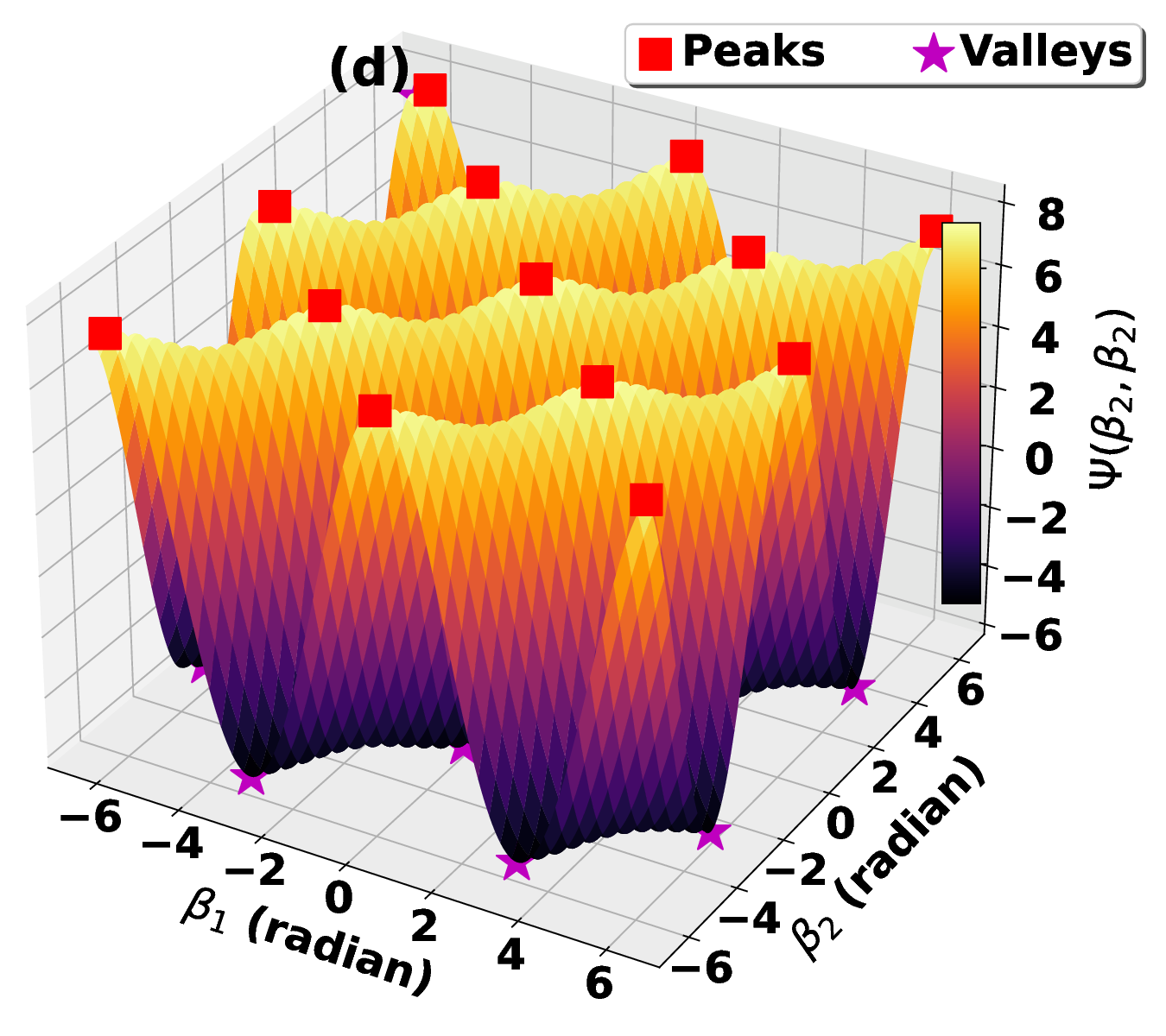}
    \caption{Surface of $\Psi(\beta_1, \beta_2)$ for the mixed frequency input 
laser pulse (M-pulse) versus phases ($\beta_1,\beta_2$) with $\omega_{12} = 
0.1, 0.2, 
0.3, 0.9$ [(a)-(d)], and $\omega_2$ for $800$~nm. Only a 
few ($\beta_1,\beta_2$) combinations may maximize/minimize $\Psi$ in Eq.\eqref{eq23}.}
    \label{Figure2_label}
\end{figure}
Figures~\ref{Figure2_label}(a)-(d) show the respective $\Psi(\beta_1, 
\beta_2)$ surface for 
$\omega_{12} = 0.1, 0.2, 0.3, 0.9$. It is clear that $\Psi(\beta_1, \beta_2)$ 
assumes different values (including maximum/minimum) depending upon combinations 
of phases $\beta_1, \beta_2$, leading to different PF {\emph{even if}} 
$\omega_{12}$ remains the same.

\begin{figure}[hbt!]
        \centering
\includegraphics[width=0.485\textwidth,height=0.995\textwidth]{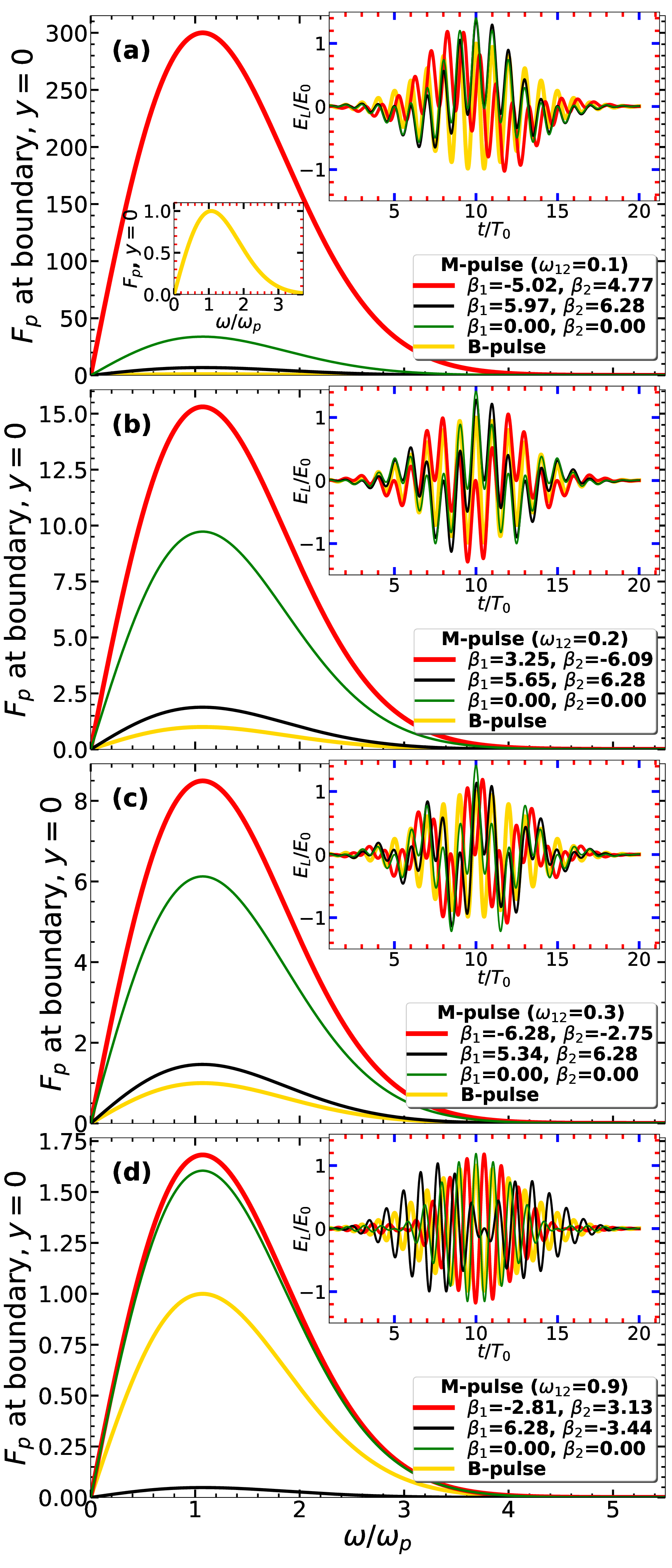}
\vspace{-0.5cm}
    \caption{Normalized PF [$F_p(\omega)$] at the plasma 
boundary~$y\!\!=\!\!0$ versus normalized frequency $\omega/\omega_p$: (a)-(d) 
show PF 
generated  by mixed frequency input laser pulse (M-pulse) with $\omega_{12} = 
0.1, 0.2, 0.3, 0.9$  respectively for different combinations of ($\beta_1, 
\beta_2$). Each panel depicts 
maximum, minimum, and an intermediate case of PF for the M-pulse along with 
the PF for the bipolar input laser pulse (B-pulse) of same energy and frequency 
$\omega_2$. The inset for $F_p$ [in (a)] shows the PF for the B-pulse  
which is more clear in (c)-(d). Other insets [(a)-(d)] show the corresponding 
electric field $E_L/E_0$ profiles for M-pulses and B-pulse. Even if 
the frequency 
$\omega_1$ remains same, the $E_L/E_0$ profiles and corresponding PFs can be 
different for different phase combinations of ($\beta_1, \beta_2$) for the 
M-pulses. }
    \label{Figure3_label}
\end{figure}

In each instance of $\omega_1$ (or $\omega_{12}$), respective normalized PF 
[$F_p(\omega, y=0)$] versus 
$\omega/\omega_p$ (in the frequency domain) are plotted in 
Fig.\ref{Figure3_label} only for those 
sets of ($\beta_1, \beta_2$) corresponding to the maximum, minimum and an 
intermediate PF case for M-pulse alongside the corresponding B-pulse case. The 
resulting M-pulse electric fields  $E_L/E_0$ versus time $t/T_0$ (after the 
$\Psi$ optimization) corresponding to these PF cases are shown (in insets) 
along with the 
B-pulse of same energy.  Normalization of PF is done w.r.t. to the 
peak of the PF 
of the B-pulse. 
%
As an example, with $\omega_{12}\!\! = \! 0.1$ in 
Fig.\ref{Figure3_label}(a), normalized 
PF shows a peak maximum (minimum) value at $\approx{293 (7)}$ near
$\omega_p$, compared to the peak value of {\emph{unity}} of
the normalized PF for
the B-pulse. An almost 300 times increase in maximum PF compared to the B-pulse case, is 
{\emph{particularly notable}}. With $\omega_{12} = 
0.2, 0.3$ in Figs.\ref{Figure3_label}(b),(c); respective 
maximum PF values for the M-pulses are noted to be 15 and 8 times higher than 
the peak of the 
B-pulse. Results clearly show
role of phases $\beta_1, \beta_2$ of
the constituent pulses of M-pulse to optimize $\Psi$ and hence PF, PP. 

Figure~\ref{Figure3_label} also show that for $\omega_1\ll\omega_2$, the 
maximum of PF for a M-pulse becomes more enhanced compared to the case of the 
B-pulse of frequency $\omega_2$ for a particular $\beta_1,\beta_2$ combinations. 
This may be understood due to dominant contribution of the terms of 
$a_1^2/\omega_{12}^2$ compared to the terms of $a_1 a_2/\omega_{12}(1 - 
\omega_{12})$ in~\eqref{eq23}. In the other regime of higher 
$\omega_1\rightarrow \omega_2$, opposite may happen, particularly when $\omega_1 
> 0.5\omega_2$. The latter case is shown in Fig.~\ref{Figure3_label}(d) 
for $\omega_{12}=0.9$. Here, the frequency difference being less, both 
the 
constituent pulses of the M-pulse may act almost similarly 
(constructively/destructively) and PF for the M-pulses for some 
$\beta_1,\beta_2$ combinations become comparable (constructive case) to that of 
the B-pulse. A poorer (destructive $\beta_1,\beta_2$) case is also identified 
with minimum PF peak value $(\ll 1)$ by the M-pulse much below the B-pulse 
case. 
At this point one may further argue that the PF/PP of a single B-pulse of 
frequency $\omega_1=0.1\omega_2$ alone, is nearly 100 times stronger than the 
PF/PP of a B-pulse of frequency $\omega_2$ as evident from~\eqref{eq26}. 
However, 
Fig.~\ref{Figure3_label}(a) in this case show PF {\emph {is still}} nearly 
300 times stronger for certain constructive combination of phases 
$\beta_1,\beta_2$ of the M-pulse.

The choice of $\omega_{12} = 0.1, 0.2, 0.3$ (or others) 
may be envisaged from the 800~nm pulse and its harmonics 
from the same laser. One may also combine lights from two different laser 
sources, e.g., Ti-Sapphire and CO$_2$ lasers which may be little difficult and uncommon, but
possible. Importantly, if the frequency difference is more, higher is the 
effect on PF/PP as evident from
Figs.\ref{Figure3_label}(a)-(c) with proper set of ($\beta_1,\beta_2$).
The other regime of $\omega_{12}\geq 0.9$ as in
Fig.\ref{Figure3_label}(d) may also be tried where terms of $a_1 
a_2/(1 - \omega_{12})$ in~\eqref{eq23} may take the dominant role (the beating 
regime). In this case $\omega_1, \omega_2$ are very close, and may be chosen 
from the same lasers.
%

%

The resulting electric fields for the M-pulses (insets) 
respective to those PF cases in Figs.\ref{Figure3_label}(a)-(d) show the 
role of phases ($\beta_1, \beta_2$) for varying electric fields even for a 
fixed $\omega_{12}$, leading to different PF/PP. The role of ($\beta_1, 
\beta_2$) showing 
beating of the constituent pulses (with 
constructive/destructive interference) of the M-pulse for $\omega_{12}\geq 
0.9$ in
Fig.\ref{Figure3_label}(d) are also very clear and explains much 
weaker PF$\ll 1$ (destructive case) than the B-pulse case.
{{Our findings on $\Psi$ \eqref{eq23} clearly connects PF/PP from low to high
frequency ratio $\omega_{12}$.}}
\vspace{-0.25cm}
\section{Terahertz field and its properties}
\label{sec4}
\vspace{-0.25cm}
Upon substituting the respective expressions \eqref{eq27a}, \eqref{eq29} of PP 
in \eqref{eq12}-\eqref{eq13}, one obtains the generated magnetic field for the 
M-pulse and 
B-pulse, denoted as $B_{zM}$ and $B_{zB}$ respectively
\begin{align}
&\quad\hspace{-3.75cm}
B_{zM}(\bold{r}, \omega) = \frac{\Psi}{\vert T_2(\omega_2,\alpha) \vert^2} \,\,\,
B_{zB} (\bold{r}, \omega)
\label{eq31}
\end{align}
\vspace{-0.5cm}
\begin{align}
&\quad\hspace{-0.75cm}
B_{zB}(\bold{r}, \omega) = \frac{i\omega_p^2\sin{\alpha}}{c\omega}
\left(\frac{\sqrt{\pi}\tau \phi_0 
|T_2(\omega_2,\alpha)|^2}{\epsilon(\omega)\cos{\alpha}
+ \sqrt{\epsilon(\omega)-\sin^2{\alpha}}}\right) \nonumber \\
&\quad \hspace{-0.85cm} \times
\exp{\left(-{\omega^2\tau^2}/{4}\right)} \exp{\left(i\omega y''/c
+ i\omega t_0 \right)}, \,\,\,\, y\leq 0.
\label{eq32}
\end{align}
Equations \eqref{eq31} and \eqref{eq32} show that low-frequency ($\omega$) wave generated at the plasma-boundary $y=0$ propagates
into the vacuum along the reflected laser pulse direction defined
by $y''=x\sin{\alpha}-y\cos{\alpha}$. Also, $B_{zM}/B_{zB} = \phi_M/\phi_B = 
\Psi/T_2^2$ holds. Similarly, fields can be obtained inside the plasma $y\ge 
0$ also. 


The energy density of the generated radiation per unit frequency is 
evaluated as
$
{dW}/{d\omega} = ({c}/{4\pi^2})|B_z(\bold{r}, \omega)|^2
$
which can be restated as
\begin{equation}
\begin{aligned}
\frac{dW}{d\Omega} = \frac{V_0^2}{c^2}\frac{\mathcal{E}_L}{\pi}\frac{dw(\Omega)}{d\Omega}
\end{aligned}
\label{eq34}
\end{equation}
where $\mathcal{E}_L = \sqrt{\pi}\tau I_L$ is the energy-density with 
intensity $I_L =
{cE_{0}^2}/{8\pi}$ of the laser pulse, $V_0 =
{eE_{0}}/{m_e\omega_2}$ is the quiver velocity of an electron, $\Omega
={\omega}/{\omega_p}$, $N={\omega_p^2}/{\omega_2^2}$ are the
normalized frequency, and electron density respectively.
The normalized energy-density per unit frequency ${dw}/{d\Omega}$ [using 
\eqref{eq31} \eqref{eq32} in \eqref{eq34}] of the generated
radiation for the M-pulse can be written in terms of B-pulse as
%
%
\begin{equation}
\begin{aligned}
& \quad\hspace{-4.6cm} \frac{dw^M}{d\Omega} =  \frac{\Psi^2}{|T_2(\omega_2,\alpha)|^4}
\frac{dw^B}{d\Omega}
\end{aligned}
\label{eq35}
\end{equation}
\vspace{-0.5cm}
\begin{equation}
\begin{aligned}
&\quad \hspace{-2.25cm}
\frac{dw^B}{d\Omega} = 
 \frac{\sqrt{\pi}\omega_2\tau}{8}N^{\frac{3}{2}} 
|T_2(\omega_2,\alpha)|^4 \\
&\quad \hspace{-1.5cm} \times \frac{\Omega^2\sin^2{\alpha}
\exp{\left(-{{N \Omega^2}{\omega_2}^2\tau^2}/{2} \right)}}  
{|(\Omega^2-1)\cos{\alpha}+\Omega\sqrt{\Omega^2\cos^2{\alpha}-1}|^2}.
\end{aligned}
\label{eq36}
\end{equation}
\begin{figure}
    \centering
\includegraphics[width=0.485\textwidth]{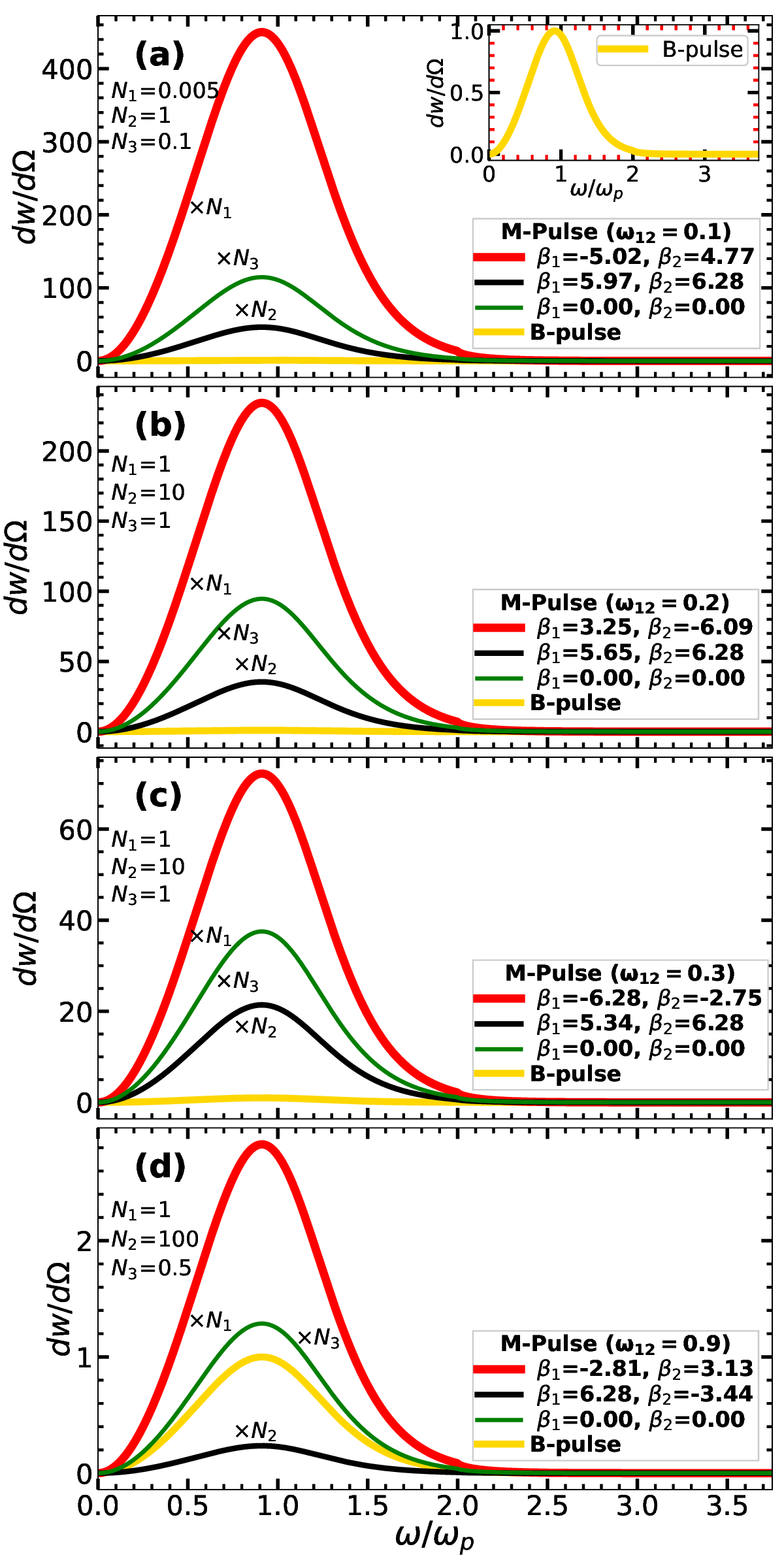}
\vspace{-0.5cm}
    \caption{Normalized radiated energy-density per unit frequency 
($dw/d\Omega$, radiation 
spectrum) at $y<0$ versus normalized frequency $\omega/\omega_p$: 
(a)-(d) show 
generated low-frequency terahertz radiation by mixed frequency input laser pulse 
(M-pulse) with $\omega_{12} = 0.1, 0, 0.3, 0.9$  respectively for various 
combinations of ($\beta_1, \beta_2$) corresponding to those parameters in 
Figs.\ref{Figure3_label}(a)-(d). Each panel shows maximum, minimum, and 
an intermediate case of radiation-spectrum for the M-pulse. Results in each 
case of $\omega_{12}$ are multiplied by different numbers $N_1, N_2, N_3$ for 
better 
visibility. The inset [in (a)] shows the radiation spectrum 
for the B-pulse with frequency $\omega_2$ and same energy of a M-pulse. Even 
if the frequency 
$\omega_1$ remains same, strength of LF terahertz radiation can be different for 
different phase combinations of ($\beta_1, \beta_2$) as respective PF are 
different.}
    \label{Figure4_label}
    \vspace{-0.2cm}
\end{figure}

This shows strength of the generated THz radiation spectrum for the M-pulse is 
$\Psi^2$ times the strength corresponding to the B-pulse. We investigate it at 
the vacuum-plasma interface $y=0$, assuming a constant $\omega_p = 
0.07\,\omega_2$ ($\nu_p =\omega_p/2\pi \approx 26.25$ THz) and an 
incident angle $\alpha = 
\pi/4$ with four values of $\omega_{12} = 0.1, 
0.2, 0.3, 0.9$ as in respective Figs.\ref{Figure3_label}(a)-(d). 
Accompanying Figures~\ref{Figure4_label}(a)-(d) illustrate 
equations \eqref{eq35} and \eqref{eq36} [inset plot], 
depicting the THz 
radiation produced by the M-pulse and B-pulse respectively, corresponding to
those judicious choice of parameters and phases $\beta_1, \beta_2$ in respective
Figs.\ref{Figure3_label}(a)-(d).
The spectral profile of the THz radiation in Fig.\ref{Figure4_label} is 
analogous for a M-pulse and B-pulse,
characterized by a Gaussian envelope (broadband) centered near $\omega_p$ 
with a spectral-width inversely 
proportional to pulse duration (shown later). Both spectra 
exhibit 
similar dependencies on the
incident angle, frequency, plasma density, and transmission coefficient, with 
the primary distinction being in the much stronger peak (for $\Psi^2$ 
dependence) in case of M-pulse compared to the case of B-pulse in most cases, 
except the near perfect destructive interference case in 
Fig.\ref{Figure4_label}(d) for $\omega_{12}=0.9$ in the beating regime 
where ponderomotive force is also very weak [Fig.\ref{Figure3_label}(d)].
%
For each $\omega_1$, optimized PF 
[respective Figs.\ref{Figure3_label}(a)-(d)] 
increases the peak of the radiation spectrum $dw/d\Omega$ with M-pulse 
(e,g, for $w_{12}=0.1$ it increases to $\approx$ {85876, 
198, 70, and 3}
times 
the B-pulse, respectively) due to $\Psi^2$ dependence. Note that for a 
single B-pulse of low frequency $\omega_1 = 0.1\omega_2$, 
the maximum 
expected strength of $dw/d\Omega$ is {\emph{only 100 times}} the B-pulse 
of frequency 
$\omega_2$. But, for the M-pulse with $\omega_{12}=0.1$ it is almost 
$8.59\times 10^{4}$ times the B-pulse case which is {\emph{very significant}}.

Thus spectral intensity of the emitted radiation is greatly enhanced with 
judicious choice of phases ($\beta_1, \beta_2$) for the M-pulse, even 
with the same low-frequency value $\omega_1$.
%
Additionally, reducing the
low-frequency value leads to a more pronounced output, provided ($\beta_1, 
\beta_2$) pairs are
carefully adjusted. The spectral peak is close to $\omega_p$. When $\omega_p
\tau \approx 1$, a wake field is generated in the plasma behind the laser pulse, 
which emits radiation around $\omega_p$ at the boundary.

\vspace{-0.5cm}
\subsection{Space-time variation of the generated magnetic field}
\begin{figure}[]
    \centering
\includegraphics[width=0.45\textwidth]{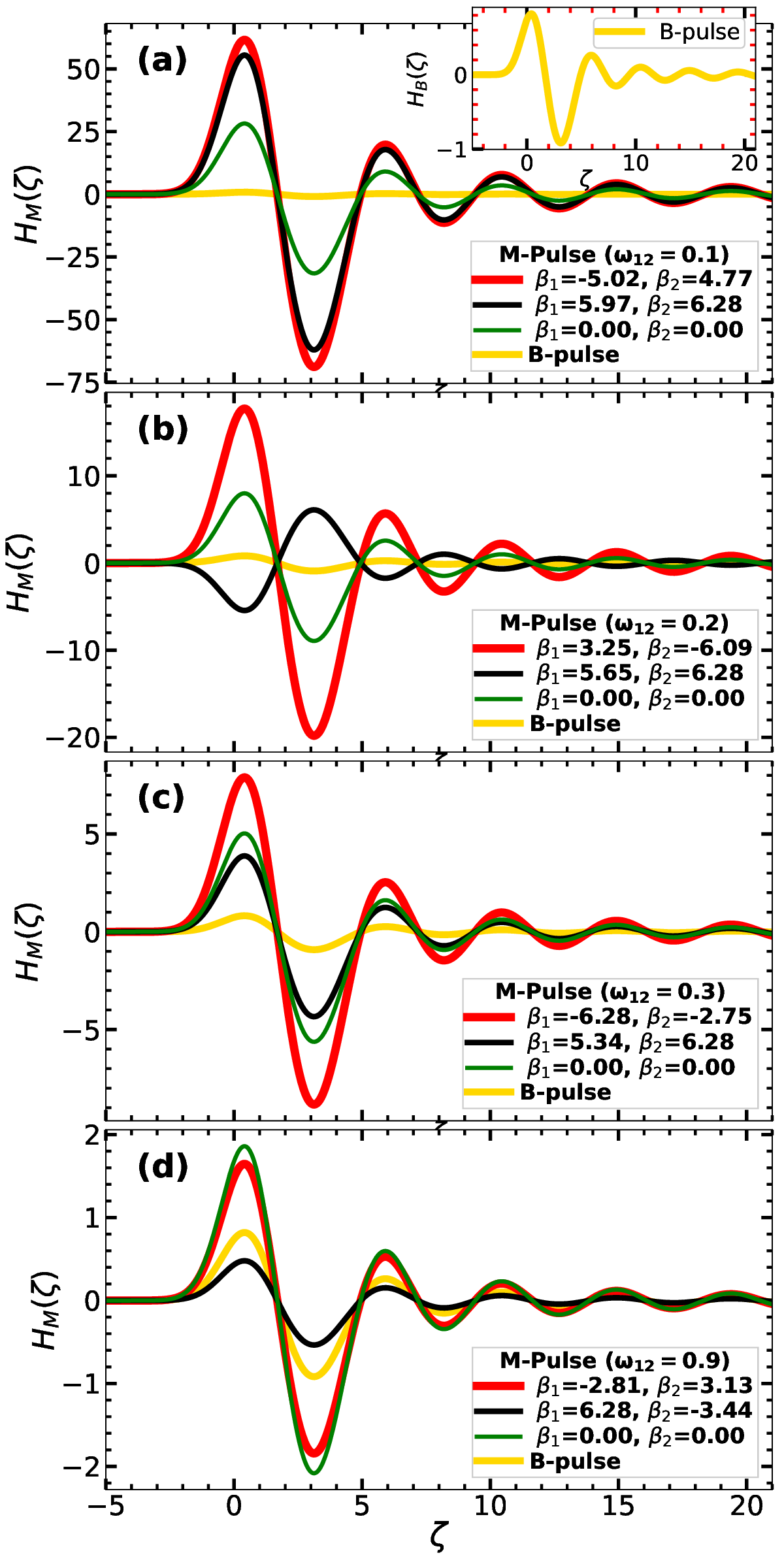}
\vspace{-0.35cm}	
	\caption{Spatio-temporal variation of the generated THz magnetic field for
$\omega_{12} = 0.1, 0.2, 0.3, 0,9 \omega_2$ [(a)-(d)] respectively, 
corresponding to those radiation-spectrum generated by M-pulses for different 
combinations
of ($\beta_1, \beta_2$) in Figs.\ref{Figure4_label}[(a)-(d)]. The inset 
[in (a)] also clearly shows the result 
for the bipolar input laser.}
    \label{Figure5_label}
    \vspace{-0.5cm}
\end{figure}

\vspace{-0.25cm}
To examine the space-time variation of the generated THz magnetic
field, we inverse Fourier 
transform~\eqref{eq31}-\eqref{eq32}. Writing $\zeta = \omega_p(t - y''/c)$,
 the generated THz magnetic fields for the M-pulse and B-pulse, in vacuum are 
respectively
%
%
\begin{equation}
\begin{aligned}
B_{zM}(\bold{r},t) =
\left({\omega_p\phi_0 \sin{\alpha}}/{\sqrt{\pi} c }\right) H_M(\zeta), 
\hspace{1cm} y \leq 0
\end{aligned}
\label{eq37}
\end{equation}
\vspace{-0.5cm}
\begin{equation}
\begin{aligned}
B_{zB}(\bold{r},t) =
\left({\omega_p\phi_0 \sin{\alpha}}/{\sqrt{\pi} c }\right) H_B(\zeta), 
\hspace{1cm} y \leq 0
\end{aligned}
\label{eq37a}
\end{equation}
%
%
%
where
\vspace{-0.25cm}
\begin{equation}
\begin{aligned}
%
& \quad \hspace{-20.0pt} H_M(\zeta) = \frac{\Psi}{|T_2(\omega_2,\alpha)|^2} \, 
H_B(\zeta), \hspace{2.25cm} y \leq 0
\vspace{-0.5cm}
\end{aligned}
\label{eq38}
\end{equation}
\vspace{-0.5cm}
\begin{equation}
\begin{aligned}
&\quad \hspace{-15.0pt}
H_B(\zeta) =  |T_2(\omega_2,\alpha)|^2 \omega_p \tau  \\
&\quad \hspace{-0.5cm} \times \mathrm{Re}\int_{0}^\infty \!\! d\Omega
\frac{\Omega\exp{\left(\!\! -i\Omega\zeta - {\omega_p^2\tau^2\Omega^2}/{4} + i\pi/2  \right)}}
{\left(\Omega^2 -1 \right)\cos{\alpha}+\Omega\sqrt{\Omega^2\cos^2{\alpha}-1}}, \hspace{0.1cm} y \leq 0.
\end{aligned}
\label{eq41}
\end{equation}
%
%
%
The functions $H_M(\zeta)$ and $H_B(\zeta)$ measuring the corresponding 
spatial-temporal distribution of
magnetic fields for M-pulse and B-pulse versus
$\zeta$ are depicted  in Figures
\ref{Figure5_label}(a)-(d) for the same parameters of 
Figs.\ref{Figure4_label}(a)-(d) with the respective combination of phases 
($\beta_1, \beta_2$).
%
%
%
%
Results are normalized by $\max[H_B(\zeta)]$, and consistent with the 
radiation spectrum in 
Figs.\ref{Figure4_label}(a)-(d), for
$\omega_{12}\!\!  =\!\!  0.1, 0.2, 0.3, 0.9$ and respective sets of phases 
($\beta_1, \beta_2$) fixed for each scenario. It is noted that peak
of $H_M(\zeta)$ for $\omega_{12} = 0.1$ are 
nearly {65} times higher than the B-pulse case. Similarly, for 
$\omega_{12} =
0.2, 0.3, 0.9$; respective $H_M(\zeta)$ maxima are nearly {18, 8, 2} 
times higher. 
A 
single B-pulse of low frequency $\omega_1\!\! = \!\!0.1\omega_2$, can only 
yield the 
maximum strength of $H_{B}(\zeta)\!\! \approx\!\! 10$, compared to the B-pulse 
of 
frequency~$\omega_2$. Thus a phase-optimized M-pulse with $\omega_{12}=0.1$ 
yields almost 
two (one) order higher output field strength, than due to a single 
B-pulse of frequency $\omega_2$ ($\omega_1$) of same energy.

Under the above conditions, the THz pulse duration 
as found in Figure~\ref{Figure5_label} is
significantly longer than that of the input laser pulse. This phenomenon 
results 
from the efficient excitation of long-lasting wake plasma oscillations within 
the plasma volume when the input laser pulse is sufficiently short. These 
oscillations are converted from the electrostatic to electromagnetic mode near 
the plasma
boundary and are subsequently emitted into the vacuum.

\vspace{-0.45cm}
\section{Phase-controllable Terahertz radiation with different pulse duration}
\label{sec5}
\begin{figure}[hbt!]
    \centering
\includegraphics[width=0.475\textwidth]{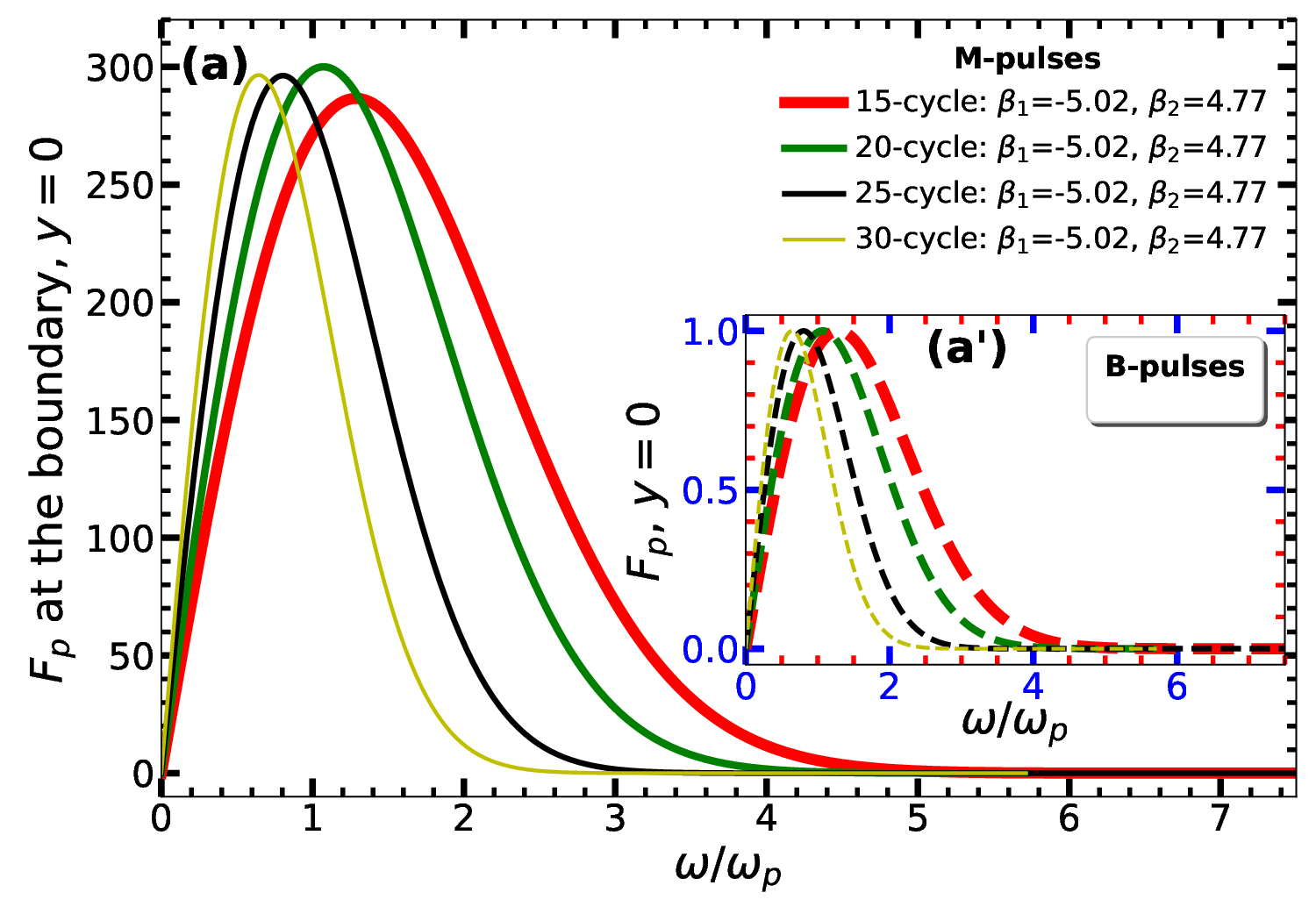}
\includegraphics[width=0.475\textwidth]{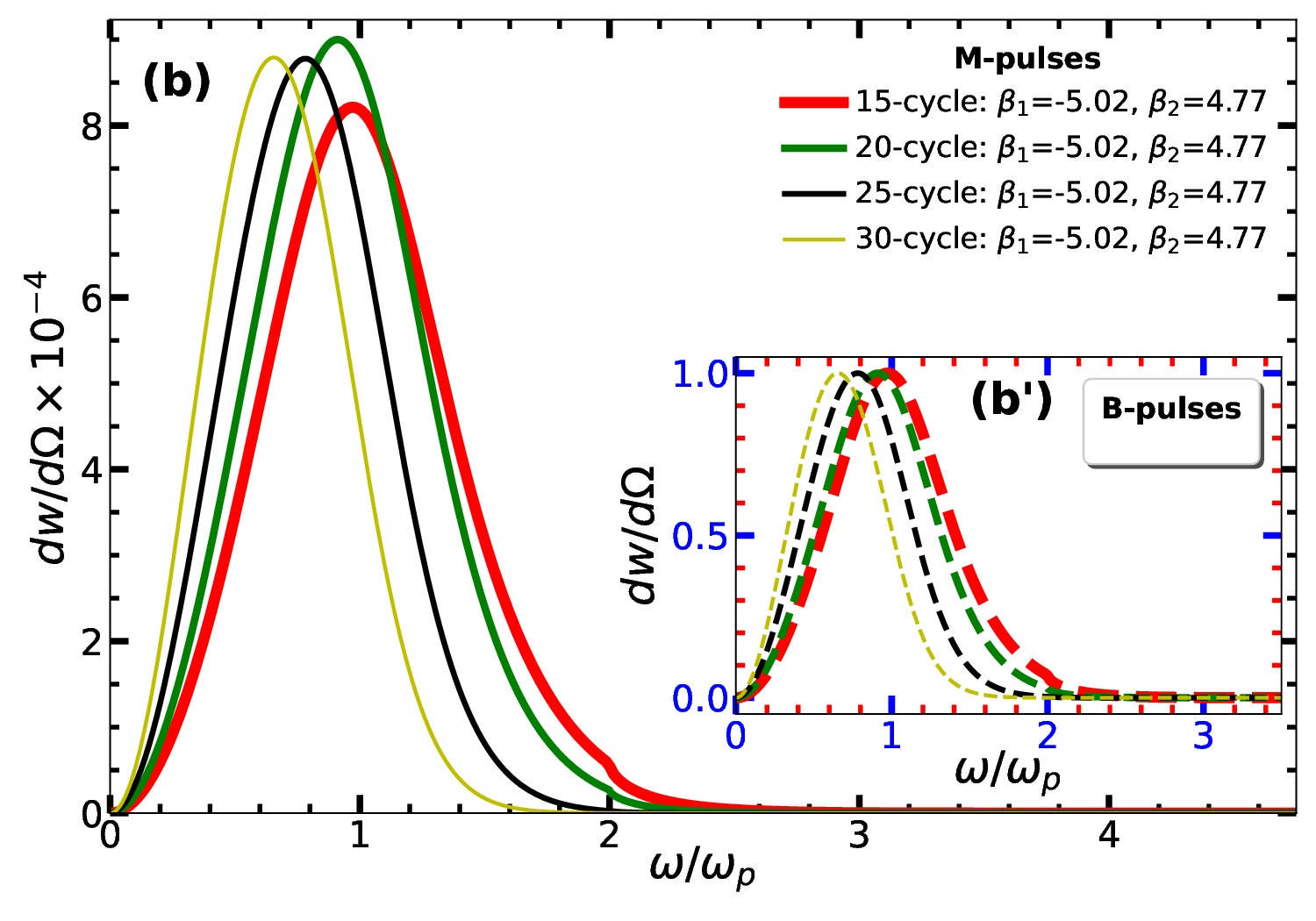}
\includegraphics[width=0.475\textwidth]{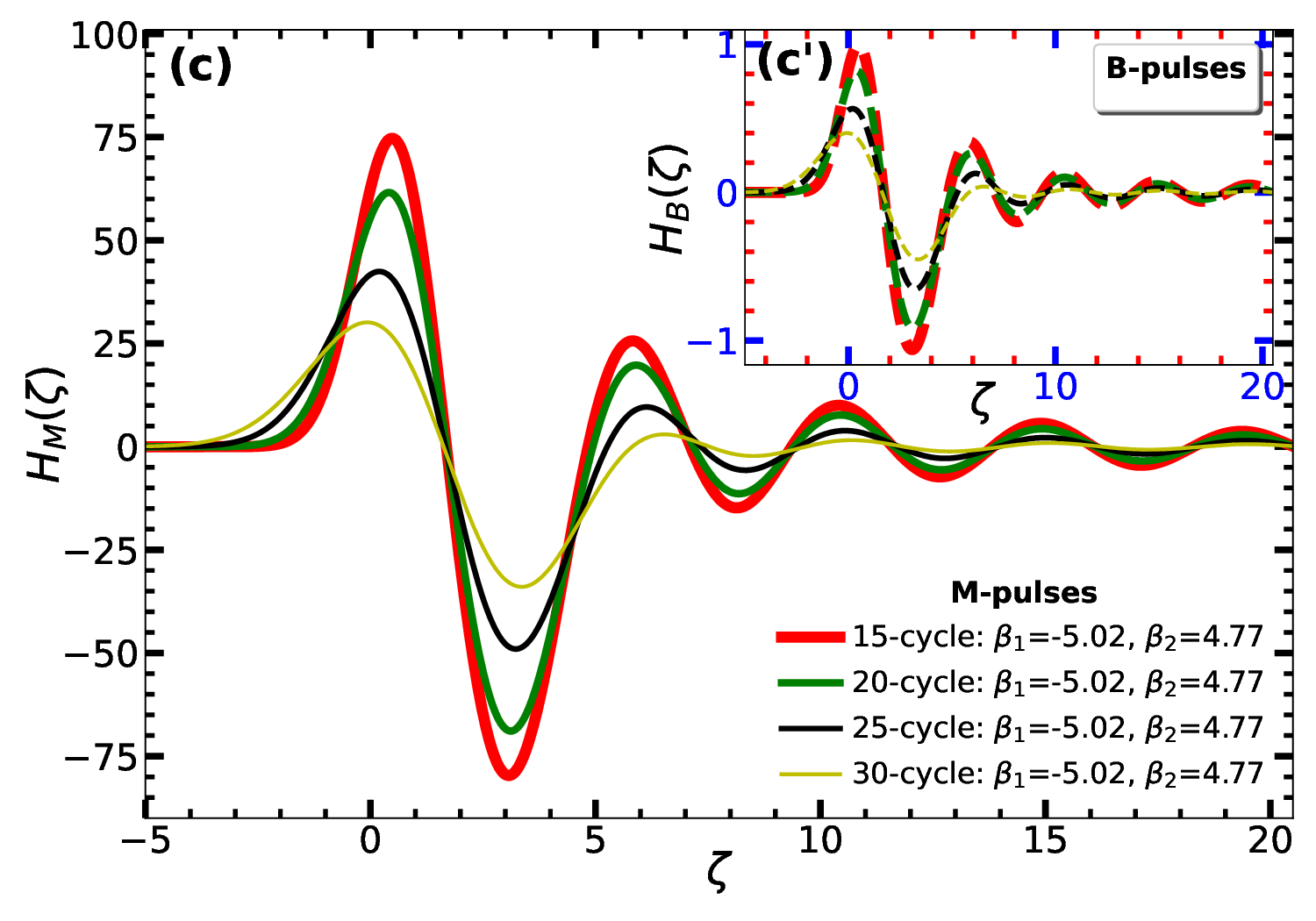}
\vspace{-0.45cm}
	\caption{\hspace{-0.25cm}(a) Normalized PF versus frequency 
$\omega/\omega_p$ at the plasma boundary $y=0$, (b) Normalized radiated energy 
per unit frequency ($dw/d\omega$, radiation spectrum) at $y<0$ versus frequency 
$\omega/\omega_p$, and (c) Generated normalized magnetic field $H(\zeta)$ versus 
$\zeta = \omega_p(t - y/c)$ at $y<0$ for different pulse duration $T = 15T_0, 
20T_0, 25T_0, 30 T_0$ with a fixed frequency ratio $\omega_1 = 0.1 \omega_2$, 
phase combination $\beta_1 = -5.02, \beta_2=4.77$ corresponding to the maximised 
$\Psi$ where effect is found to maximum (e.g., 
Figs.\ref{Figure3_label},\ref{Figure4_label},
\ref{Figure5_label}). The insets [(a'), (b'), (c')] in each plot 
show 
respective results for the 
bipolar 
input laser-pulses of same energy for same parameters with respective M-pulses.}
    \label{Figure6_label}
    \vspace{-0.5cm}
\end{figure}
\vspace{-0.25cm}
It is now important to understand how the generated terahertz radiation 
spectral-width ($\Delta\omega$ at FWHM) and the associated THz field change 
with the laser pulse duration~$T$. To illustrate this, we choose M-pulses with 
a fixed $\omega_1 = 0.1 \omega_2$, phase combination $\beta_1 = -5.02, 
\beta_2=4.77$ corresponding to the maximised $\Psi$ where effect is found to 
be maximum (e.g., 
Figs.\ref{Figure3_label},\ref{Figure4_label},
\ref{Figure5_label}) and vary the pulse duration $T = 15T_0, 
20T_0, 
25T_0, 30 T_0$. The respective B-pulses with a fixed frequency $\omega_2$ have 
also same respective durations ($T$) and energies of the M-pulses. 

Figure~\ref{Figure6_label}(a) shows the normalized PF [$F_p(\omega)$] generated at 
the plasma boundary $\!y=\!0$, for different $T$. The inset plot~[(a')] shows 
respective results for the input B-pulses. It is seen that peak
strengths of PFs of M-pulses are maintained almost (285-300) times higher 
compared to that (unity) of respective B-pulses. However, their width 
in the frequency domain increases with shortening of pulse duration for 
both M-pulses and B-pulses. Thus, a shorter pulse is important to generate 
wide-band PF. This effect of PF is clearly shown in the respective 
radiation spectrum ($y<0$) in Fig.~\ref{Figure6_label}(b) for M-pulses and 
Fig.~\ref{Figure6_label}(b') for B-pulses. The peak strengths of radiation spectrum 
($dw/d\Omega$) of M-pulses are nearly $(8-9)\times 10^4$ times higher compared 
to the peak radiation strengths (unity) of respective B-pulses. Along with the 
gradual widening of the radiation spectrum due to successive pulse-shortening 
a substantial systematic blue-shift (a shift toward high-frequency) w.r.t. the 
plasma frequency $\omega_p$ is noted that corroborates PF results in 
Figs.~\ref{Figure6_label}(a),6(a'). This particular findings are important, for 
frequency selection of generated THz. The corresponding normalized magnetic 
fields $H(\zeta)$ versus $\zeta = \omega_p(t - y''/c)$ far from the plasma 
surface ($y''<0$) in Figure~\ref{Figure6_label}(c) for M-pulses and 
Figure~\ref{Figure6_label}(c') for B-pulses show much stronger THz field for M-pulses 
with shorter pulses, e.g., $T=15 T_0$.

\begin{figure}[hbt]
    \centering
\includegraphics[width=0.4\textwidth]{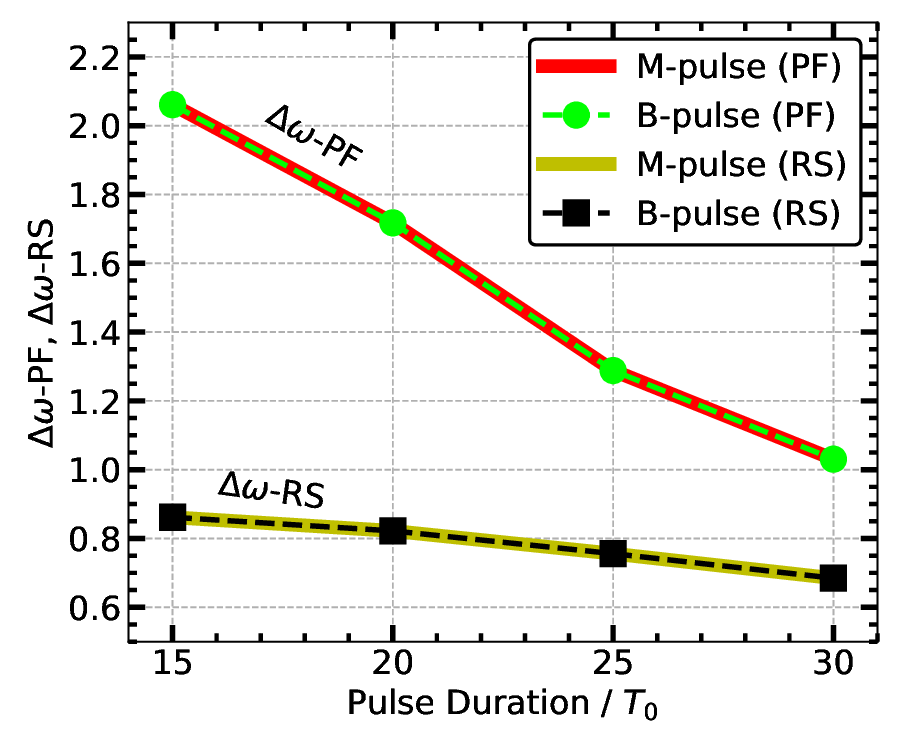}
\vspace{-0.25cm}
	\caption{Spectral-width ($\Delta\omega$ at FWHM) of PF [from 
Fig.\ref{Figure6_label}~(a),(a')] and corresponding spectral width of radiation 
spectrum versus the laser pulse duration $T/T_0$ for same parameters with 
respective M-pulses and B-pulses as in Figs.\ref{Figure6_label}. Other parameters are 
as in Fig.\ref{Figure3_label}
}
    \label{Figure7_label}
\end{figure}

We calculate the spectral-width ($\Delta\omega$ at FWHM) for PFs [from 
Figs.\ref{Figure6_label}~(a),(a')] and radiation spectrum (RS) [from 
Figs.\ref{Figure6_label}~(b),(b')] with different $T$ for 
respective M-pulses and B-pulses.
The scaling of PF spectral-width ($\Delta\omega$-PF) and the  radiation 
spectral-width ($\Delta\omega$-RS) versus $T/T_0$ is shown in 
Figure~\ref{Figure7_label}. Both the spectral-widths are found to be inversely 
proportional to the pulse duration. The narrow spectral-width of the 
RS-spectrum compare to that of the PF-spectrum can be qualitatively 
understood from their respective terms of
$\sim \exp(-\tau^2 \omega^2/4)$ and $\sim \exp(-\tau^2 \omega^2/2)$ dependence.

\vspace{-0.45cm}
\section{Summary}
\label{sec6}
\vspace{-0.25cm}
We have investigated terahertz (THz) radiation generation at the vacuum--plasma 
interface driven by the oblique incidence of s-polarized Gaussian laser 
pulse(s) on an underdense plasma, with a focus on enhancing emission through 
two-color mixed-frequency excitation schemes (M-pulses). Moving beyond the 
conventional single-frequency bipolar (B-pulse) approach, our study reveals that 
carefully tailored M-pulses -- comprising two frequency components with 
adjustable amplitude and phase -- can break the temporal symmetry of the 
effective laser field and induce a significantly stronger ponderomotive force 
(PF), which serves as the main driver for THz emission via TR.

A new analytical expression for the PF has been derived, which incorporates the 
phase asymmetry effects unique to M-pulses. This formulation explains the 
enhanced THz generation as a direct consequence of cycle-to-cycle symmetry 
breaking in the driving field. Notably, we show that M-pulses with the same 
total energy as B-pulses can produce orders-of-magnitude stronger PF and 
resulting THz radiation spectrum (RS). By optimizing the frequency ratio and 
phase difference 
between the pulse components, a locally unipolar (in this work) or globally 
unipolar-like (which will be discussed in detail in our forth coming article) 
field structures can be formed, offering efficient and tunable THz emission. 

With the gradual broadening of the spectral-width $(\Delta\omega)$ 
of PF and RS due to successive pulse-shortening a substantial systematic 
blue-shift in their respective peak values
w.r.t. the plasma frequency have been identified.
These findings suggest that phase-engineered two-color M-pulses are a powerful 
and flexible tool for driving laser--plasma-based THz sources, with 
promising implications for the design of compact, high-efficiency broadband THz 
systems.

The phase-dependent enhanced PF
generated by M-pulses suggests potential applications in various
fields, including enhanced laser-driven wakefield acceleration, ion-acceleration, inertial confinement fusion, and high harmonic generation.


\bibliographystyle{unsrt}
\bibliography{Biblography}

\end{document}